# Superradiance from Lattice-Confined Atoms inside Hollow Core Fibre

Shoichi Okaba[1,2], Deshui Yu[1], Luca Vincetti[3], Fetah Benabid[4,5] & *Hidetoshi Katori[1,2,6]

**Unravelling superradiance, also known as superfluorescence, relies on an ensemble of phase-matched dipole oscillators and the suppression of inhomogeneous broadening. Here we report on a novel superradiance platform that combines an optical lattice free from the ac Stark shift and a hollow-core photonic crystal fibre, enabling an extended atom-light interaction over 2 mm free from the Doppler effect. This system allows controlling the atom spatial distribution and spectral homogeneity whilst efficiently coupling the radiation field to an optical fibre. The experimentally-observed and theoretically-corroborated temporal, spectral and spatial dynamic behaviours of the superradiance, e.g., superradiance ringing and density-dependent frequency shift, demonstrate a unique interplay between the trapped atoms and the fibre-guided field with multiple transverse modes. Our theory indicates that the resulting temporal evolution of the guided light shows a minimal beam radius of 3.1 μm which is three times smaller than that of the lowest-loss fibre mode.**

Superradiance (SR) has been the subject of active research since the pioneering work of Dicke in 1954[1]. Such a collective effect arises from an ensemble of two-level emitters spontaneously locking their dipole oscillations in phase[2], giving rise to an enhanced decay rate $\gamma_{SR}$ multiple times faster than the spontaneous emission rate $\gamma_0$ of individual emitters and a strong peak intensity of radiation in quadratic proportion to the excited-state population. The other SR signature manifests as a collective frequency shift and broadening, which are underpinned by fascinating physical phenomena such as the energy-level shift due to virtual photon exchange (collective Lamb shift)[3] or van der Waals dephasing[2]. In addition to its undisputed importance in fundamental science, SR also has substantial applications in various fields. For example, lasing based on SR has been proved to be a potentially outstanding frequency reference because of its strong immunity to environmental perturbations[4,5]. In quantum information, SR is utilized in realizing quantum memories[6,7] and single photon sources[8]. Also, SR has been recently used for cooling mechanisms by means of cooperative emission[9,10].

[1] Department of Applied Physics, Graduate School of Engineering, The University of Tokyo, 7-3-1 Bunkyo-ku, Tokyo 113-8656, Japan. [2] RIKEN Center for Advanced Photonics, 2-1 Hirosawa, Wako, Saitama 351-0198, Japan. [3] Department of Engineering "Enzo Ferrari" University of Modena and Reggio Emilia via Vivarelli 10, I-41125 Modena, Italy. [4] GPPMM group, Xlim Research Institute, CNRS UMR7252, 123, Albert Thomas, Limoges 87060, France. [5] Physics department, The University of Western Australia, Crawley 6009, WA, Australia. [6] Quantum Metrology Laboratory, RIKEN, 2-1 Hirosawa, Wako, Saitama 351-0198, Japan. Correspondence and requests for materials should be addressed to H.K. (email: katori@amo.t.u-tokyo.ac.jp).

Despite the observations of SR in various gaseous and solid-state samples[11-19], capturing the full picture of SR properties within a single experimental platform still remains an ongoing target. The chief challenge in unravelling the SR dynamics is to find the best trade-off between maximising the number of phase-locked emitters coupled to the common radiation field and preserving the inter-emitter phase correlation. The enhancement of SR emission rate scales as the former while the latter is a prerequisite for the SR generation. The conventional SR schemes adopt an ensemble of emitters densely packed within a sub-radiation-wavelength volume, i.e., Dicke regime[11-14], to achieve the SR. Recently, a large number of atoms confined inside optical cavities[4-10] or coupled to waveguides[19] have been a successful alternative for SR generation. This is based on the fact that the fraction of photon energy emitted into the SR mode $N\eta/(1 + N\eta)$ approaches unity for the atomic number $N \gg 1$ even when the single-atom cooperativity parameter $\eta = \gamma_{SR}/\gamma_0$, which is a geometric parameter characterizing the emissive coupling of an atom to the cavity mode, is much less than unity[8].

Research at the interdisciplinary frontier between the fields of cold atoms and photonic materials reports promising results in controlling ultracold atoms with photonic waveguides[20], which holds the potential of offering a strong cooperative enhancement while keeping the low atomic density. Indeed, ultracold atoms are especially attractive materials since they possess the distinct feature of being precisely controllable by optical lattice[21]. In particular, when ultracold atoms are confined in an optical lattice tuned to the magic wavelength, the optical excitation of atomic transition is not only free from the Doppler shift but also free from the lattice-field-induced ac Stark shift[22], offering an ideal platform for investigating SR based on well-isolated atomic systems. In addition, low-loss photonic waveguides allow a strongly enhanced one-dimensional (1D) atom-light interaction, as they allow the interaction length much longer than the Rayleigh range $z_R = \pi w_0^2/\lambda_0$ at the wavelength $\lambda_0$ while still keeping a small light-beam waist radius $w_0$. Successful combination of ultracold atoms and photonic waveguide has been exemplified by the works with nanofibers[23,24], photonic waveguides[19], and hollow core fibers[25]. The long-range atom-atom interaction is accomplished via guided photons of the so-called alligator photonic crystal waveguide[19], and the resulting single-atom cooperativity reaches $\eta = 0.34$. Ultracold atoms are successfully loaded inside a hollow-core photonic crystal fibre (HCPCF) and trapped in a 1D magic-wavelength optical lattice[25], revealing 10-kHz-wide linewidth of the transition without being affected by wall-atom collisions.

In the present work, we report on the observation of SR from ultracold atoms confined in the magic optical lattice inside the HCPCF[26]. We couple an HCPCF-transmitted SR to a single-mode fibre to measure the signal by employing a highly sensitive heterodyne technique, which allows investigating temporal, spectral and spatial dynamic behaviours of SR with a single shot measurement. We observe a ringing and accelerated decay that is up to 55 times faster than the spontaneous emission of single atoms. By analysing the inphase/quadrature components of the signal, we measure a density-dependent frequency shift. Finally, we discuss the multi-mode dynamics of SR in the HCPCF, where a superposition of specific guided modes contributes to the photon-mediated SR, complemented with theoretical predictions. The dense cigar-shaped atomic cloud restricts the guided SR field mainly within a diameter three times smaller than that of the fundamental fibre mode. For an in-fibre atom number of $N = 2.5 \times 10^5$ with the single-atom cooperativity $\eta = 4.1 \times 10^{-5}$ available for the HCPCF, the $N$-atom cooperativity parameter[8] $N\eta$ can be as high as 10, leading to a near-unity coupling efficiency to the guided mode of the hollow core fibre.

## Results

**Experimental scheme** Figure 1a shows the schematic of our experiment. A 32-mm-long hypocycloid-core-shaped Kagome-lattice HCPCF[26] is horizontally placed in a vacuum chamber. The inner diameter of the hollow core fibre reaches 40 μm. The optical fibre guides light via the inhibited-coupling mechanism[27] and supports multiple transverse modes whose transverse profiles are very close to those of a dielectric capillary[28]. The Gaussian-like fundamental $LP_{01}$ core mode has a $1/e^2$ waist radius $w_0 = 11.8$ μm ($1/e$ radius is 8.34 μm). $^{88}$Sr atoms are laser cooled and trapped in a two-stage magneto-optical trap (MOT)[29] at $z \approx -1$ mm (the origin of $z$-coordinates is set at the entrance end of the HCPCF) and loaded into an optical lattice which passes through the HCPCF (see Methods). The counter-propagating linearly-polarized lattice lasers match the $LP_{01}$ mode and operate at the wavelength $\lambda_L = 813$ nm. A bias magnetic field $\mathbf{B}_0 = (0.14 \text{ mT}) \hat{\mathbf{e}}_x$ is applied to define the quantization axis. We focus on the $^1S_0 -$

$^3P_1(m = 0)$ intercombination transition of $^{88}$Sr at the frequency $\omega_0 = 2\pi \times 435$ THz with the natural linewidth $\gamma_0 = 2\pi \times 7.5$ kHz and the dipole moment $D = 0.087\ ea_0$ (elementary charge $e$ and Bohr radius $a_0$). The lattice laser polarization is tuned to cancel out the ac Stark shift difference between $^1S_0$ and $^3P_1(m = 0)$ states caused by the optical lattice[21,25].

The loaded atoms are further transported to the position $z \approx 2$ mm inside the hollow-core fibre by a moving lattice with its depth of 300 μK[25]. By turning off one of the two lattice lasers, the atomic gas, which is radially guided by a travelling-wave lattice laser, spreads axially to a full width at half maximum (FWHM) $l_z$. Then, the lattice laser is switched on again to recapture the atoms. We consider two cloud widths, i.e., unexpanded ($l_z = 0.87$ mm) and expanded ($l_z = 2.1$ mm) clouds (see Fig. 1b), corresponding to the number of lattice sites $N_L = l_z/(\lambda_L/2) = 2.1 \times 10^3$ and $5.2 \times 10^3$, respectively (see Methods). Each lattice site forms a pancake-shaped harmonic potential, where the atoms are distributed within a transverse $1/e$ radius of $r_a = 1.7$ μm in the $x - y$ plane and a longitudinal $1/e$ width of $l_a = 54$ nm along the fibre axis. Varying the number of loaded atoms $N$ tunes the on-site density of atoms $\rho = N_a/(\pi r_a^2 l_a)$ with the average number of atoms in each lattice site $N_a = N/N_L$. We note that the multiple photon scattering caused by the periodic spatial distribution of atoms[30] hardly affects the atom-light interaction as the lattice period $\lambda_L/2$ does not fulfil the Bragg condition for the transition wavelength $\lambda_0$.

**Superradiance ringing** All atoms are initialized in the $^1S_0$ ground state. We apply an $\hat{\mathbf{e}}_x$-polarized light pulse $\mathbf{E}_p$ with a duration $\tau_p = 500$ ns to pump the atoms. The pump laser beam is matched to the LP$_{01}$ fibre core mode, and the in-fibre amplitude $E_p$ is set to $E_p = \hbar\pi/D\tau_p$ with $\hbar$ being the reduced Planck constant. The pump-laser frequency $\omega_p$ is stabilized to an optical cavity made of an ultralow-expansion (ULE) glass with a drift rate $\sim 1$ kHz h$^{-1}$ and detuned $\Delta_p = \omega_p - \omega_0$ from the atomic resonance $\omega_0$. The observed SR field at the fibre output end can be written as $\mathbf{E}_{SR}(\mathbf{r},t) = (\hat{\mathbf{e}}_x/2)E_{SR}(\mathbf{r},t)e^{-i\omega_0 t} + \text{c.c.}$ with the radial coordinate $\mathbf{r} = x\hat{\mathbf{e}}_x + y\hat{\mathbf{e}}_y$. We observe that the SR field propagates in the same direction as the pump light, which is consistent with the directed spontaneous emission of the timed Dicke state[3,31]. The radiation power $P_{SR}(t) \propto \iint |E_{SR}(\mathbf{r},t)|^2\ d^2\mathbf{r}$ is typically on the order of nW, requiring the low noise detection. To address this issue, we utilize the balanced-heterodyne technique[32] (see Fig. 1c).

The SR output is coupled to a single-mode fibre (SMF) and mixed with a local oscillator $\mathbf{E}_{\mathrm{LO}}(\mathbf{r}, t) = \hat{\mathbf{e}}_x E_{\mathrm{LO}}(\mathbf{r}) \cos \omega_{\mathrm{LO}} t$ by a 50:50 fibre coupler (see Methods). The power and the frequency of local oscillator are on the order of 1 mW and $\omega_{\mathrm{LO}} = \omega_0 - \Omega_0$ with $\Omega_0 = 2\pi \times 50$ MHz respectively. The relation among $\omega_0$, $\omega_{\mathrm{p}}$, and $\omega_{\mathrm{LO}}$ is summarized in Fig. 1d. Furthermore, the SMF acts as a spatial mode filter to mainly extract the $LP_{01}$ component from the SR output (see Methods) as given by $\mathbf{E}'_{\mathrm{SR}}(\mathbf{r}, t) = (\hat{\mathbf{e}}_x/2) E'_{\mathrm{SR}}(\mathbf{r}, t) e^{-i\omega_0 t} + \mathrm{c.c.}$, where a prime, such as $E'$, denotes the $LP_{01}$ component.

The heterodyne signal $V_{\mathrm{RF}}(t)$ is band-pass filtered with a bandwidth of 4 MHz to improve the signal-to-noise ratio (SNR) of the measurement (see Methods), and then is recorded by a digital storage oscilloscope (DSO). Upper panel of Fig. 2a shows a typical trace of the single-shot heterodyne signal $V_{\mathrm{RF}}(t)$, where a yellow shaded region corresponds to a 500-ns-long pump pulse. The envelope exhibits a ringing behaviour, where the first pulse-shaped envelope with a 0.7-μs width corresponds to the pumping process while the following bursts are caused by the reabsorption and reemission of radiation between different parts of atomic cloud[33,34]. The temporal widths of the bursts after the pump pulse are much shorter than the spontaneous emission lifetime $1/\gamma_0 = 21$ μs of single atom, manifesting the occurrence of SR. In particular, the first SR burst with a temporal $1/\sqrt{e}$ width of $\gamma_{\mathrm{bw}}^{-1} = 0.38$ μs gives a photon-emission enhancement factor of $\gamma_{\mathrm{bw}}/\gamma_0 = 55$. Lower panel of Fig. 2a shows a temporal variation of SR power $\propto |V_{\mathrm{RF}}(t)|^2$ in the logarithmic scale. Thanks to the heterodyne detection, we achieve the measurement dynamic range over 5 orders of magnitude. Such sensitive detection scheme will allow accessing orders of magnitude weaker subradiant process, which usually falls behind the SR process[35,36], by extending the measurement time sufficiently longer than $\gamma_0^{-1}$. Figure 2b shows the normalized burst width $\gamma_{\mathrm{bw}}/\gamma_0$ as a function of the number of atoms $N$, indicating the characteristic feature of the SR or cooperative emission that is proportional to $N$. A linear fit to the data determines the single-atom cooperativity $\eta_{\mathrm{bw}} = \frac{\gamma_{\mathrm{bw}}}{N\gamma_0} = 4.9 \times 10^{-4}$.

**Frequency shift** The signal $V_{\mathrm{RF}}(t)$ carries also frequency information. Indeed, the amplitude of SR light output from the SMF may be rewritten as $E'_{\mathrm{SR}}(\mathbf{r}, t) = A'_{\mathrm{SR}}(\mathbf{r}, t) e^{-i\Delta_{\mathrm{SR}} t}$, where the complex function $A'_{\mathrm{SR}}(\mathbf{r}, t)$ is related to the envelope of $V_{\mathrm{RF}}(t)$ and $\Delta_{\mathrm{SR}} = \omega_{\mathrm{SR}} - \omega_0$ corresponds to the shift of the centre frequency $\omega_{\mathrm{SR}}$ of the SR light from the atomic-transition frequency $\omega_0$ (Fig. 1d). The frequency shift $\Delta_{SR}$ can be extracted by I/Q demodulating the carrier signal of $V_{RF}(t)$ recorded by DSO (see Fig. 3a) into the in-phase $V_I(t)$ and quadrature-phase $V_Q(t)$ components, i.e., $V_{RF}(t) = V_I(t) \cos \Omega_0 t + V_Q(t) \sin \Omega_0 t$. Figure 3b shows an example of an amplitude $V(t) = \sqrt{V_I(t)^2 + V_Q(t)^2}$ and a phase $\theta(t) = \text{ArcTan}[V_Q(t)/V_I(t)]$ obtained from the I/Q signals, where we repeat the measurement of $V_{RF}(t)$ ten times and average the data. The linearly varying phase $\theta(t)$ in time indicates the frequency shift of the emitted light $\Delta_{SR} = d\theta/dt$, which is nearly constant over 15 μs except the discontinuities at the nodes of the amplitudes that cause $\pi$-phase jump. The same I/Q signals are parametrically mapped onto the I/Q plane as shown in Fig. 3c, where the purple line corresponds to the trajectory from $t = 1.48$ μs to 8.38 μs. By taking a certain rotating frame $\Delta_{SR} = d\theta/dt$, the trajectory can be projected to a line as indicated by the blue colour (see Methods). We thus determine $\Delta_{SR} \approx -2\pi \times 93(10)$ kHz for the unexpanded atomic cloud with $N = 9.4 \times 10^4$ ($\rho = 8.8 \times 10^{13}$ cm$^{-3}$). A slight departure of the phase $\theta(t)$ from dashed lines in Fig. 3b, which corresponds to finite width of the blue trajectory in Fig. 3c, may indicate a frequency chirp in the emitted signal. Further investigation on the chirping effect will be given elsewhere. In the following, we analyse $\Delta_{SR}$ in the latter method as illustrated in Fig. 3c, assuming $\Delta_{SR}$ to be constant.

We investigate the conditions that may affect the frequency shift $\Delta_{SR}$ of the SR field. Figure 3d shows the dependence of $\Delta_{SR}$ on the atomic density $\rho$ for the resonant pumping $\Delta_p = 0$. It is seen that $\Delta_{SR}$ is negative and $|\Delta_{SR}|$ grows up linearly with $\rho$. A linear fit finds the coefficient for the frequency shift to be $\Delta_{SR}/\rho \sim -1 \times 10^9$ Hz cm$^3$, which is consistent with the values measured in the absorption spectrum[25,37], suggesting the emission and absorption spectra are subject to the same density shift. We also measure $\Delta_{SR}$ as a function of the pump-field detuning $\Delta_p$ with the same amplitude and duration of the pump pulse as that of the $\pi$-pulse with $\Delta_p = 0$. As shown in Fig. 3e, for a given density $\rho$, $\Delta_{SR}$ stays nearly constant over a wide range of $\Delta_p$, which indicates that the shift $\Delta_{SR}$ arises entirely from the density shift and is independent of the pumping frequency $\omega_p$. For $N = 2 \times 10^5$ atoms confined in the unexpanded lattice, the mean site-occupancy is about $10^2$ and the interatomic separation is 300 nm, leading to a collective Lamb shift plus a Lorentz-Lorenz shift[38] of about $\gamma_0$. The frequency shift observed in experiment well exceeds any of them. Indeed, the resonant dipole-dipole interactions (RDDIs) between the atoms in the same lattice site primarily contribute to the observed density-dependent frequency shift. The linear dependence between $\Delta_{SR}$ and $\rho$ is well reproduced by our numerical simulation (see below and Supplementary Note 5).

**Efficiency of superradiance** We further consider the efficiency of SR that weighs the energy transfer from the pump field to the SR light. In the spontaneous emission, the emitted photons are randomly oriented in free space and thereby hardly coupled to the fibre modes. In contrast, the SR shows a well-defined direction depending upon the sample's geometry. The multi-transverse-mode propagation in the HCPCF may support an in-fibre beam which matches the atomic cloud aligned along the fibre axis. As a result, the light power output from the fibre provides the attenuated pumping pulse by the absorption of atoms and the SR field emitted from the atoms. We investigate the efficiency of SR by measuring the total radiation power $P_{SR}(t)$ for the applied pump power $P_p(t)$.

The measurement procedure is illustrated in Fig. 1a. The light output from the HCPCF is split into two paths by a beam splitter. In one arm, a photomultiplier tube (PMT) measures the total light intensity to determine the coupling efficiency $\kappa$ of the SR to multi-spatial modes while in the other arm, the beam enters the SMF. The light output from the SMF is in a superposition state of the fibre $LP_{01}$, $LP_{02}$, and $LP_{03}$ modes, where the $LP_{01}$ component dominates the weight. This light beam is detected by an avalanche photodiode (APD) so as to measure the coupling efficiency $\kappa_0$, which is mostly determined by the $LP_{01}$ mode, i.e., the efficiency of transferring the pumping power to a specific mode (see Methods). The photon counting mode is applied for both measurements, where the neutral-density (ND) filters are inserted in front of the detectors to attenuate the signal down to the photon-counting level.

Figure 4a shows a typical data obtained by the PMT after averaging 25,000 measurements with $N = 9.4 \times 10^4$. The green curve displays the result in the absence of the atoms, i.e., the power of the pump light $P_p(t)$, while the red line denotes the measurement with the presence of lattice-confined atoms, i.e., the power of the light passing through the atoms and superradiantly emitted by the atoms $P_{SR}(t)$. The blue line corresponds to the difference $\Delta P(t) = P_{SR}(t) - P_p(t)$, where the positive (negative) area is proportional to the total emitted $P_{em}$ (absorbed $P_{ab}$) power. The efficiency $\kappa$ is then given by the ratio $\kappa = |P_{em}/P_{ab}|$. Figure 4b shows the dependence of $\kappa$ on the atomic number $N$ for the unexpended (black solid squares) and expanded (red solid circles) atomic clouds. It is seen that in both cases $\kappa$ goes up monotonically and approaches unity asymptotically as $N$ is increased. The absorbed energy in the pumping process can be converted into either the SR radiation or the spontaneously emitted radiation. The former, whose power scales as $N^2 \gamma_{SR}$ with the average SR emission rate $\gamma_{SR}$ of single atom, is collectively coupled to the HCPCF with an efficiency $\chi$. In contrast, the latter is proportional to $N\gamma_0$ and hardly matches the HCPCF modes. Thus, the efficiency $\kappa$ can be formulated as $\kappa = \chi \frac{\eta N}{1+\eta N}$, where $\eta = \gamma_{SR}/\gamma_0$ denotes the single-atom cooperativity. The curve fitting to filled symbols in Fig. 4b gives $\chi \approx 1$ and $\eta \approx 4.1 \times 10^{-5}$, which reasonably agrees with the geometrically estimated value $\eta_\Omega = NA^2/4 \approx 10^{-4}$ with the numerical aperture $NA = \lambda_0/\pi w_0 \approx 0.02$ for coupling to the guided fibre mode.

Similarly, we obtain the coupling efficiency $\kappa_0$ of the superposition state of the fibre $LP_{01}$, $LP_{02}$, and $LP_{03}$ modes, as shown in Fig. 4b by open square symbols. It is found that $\kappa_0$ is maximized around $N_m \approx 10^5$ and then decreases as $N$ is further increased, suggesting that fibre higher-order modes take up more SR energy for $N > N_m$. The SR rate $\gamma_{SR}$ may be separated into two parts, which respectively contribute to the fundamental $LP_{01}$- and high-order-mode SR fields, i.e., $\gamma_{SR} = \gamma_{SR}^{(f)} + \gamma_{SR}^{(h)}$, and $\kappa_0 = \left(\gamma_{SR}^{(f)}/\gamma_{SR}\right)\kappa$. Since the atoms emit the guided light within the cross-sectional area of the atomic cloud, whose radius $r_a$ is much smaller than the $LP_{01}$-mode beam radius $w_0$, the higher-order fibre modes (such as $LP_{02}$ and $LP_{03}$) with their central-peak radii smaller than $w_0$ get enhanced prior to $LP_{01}$ and hence $\gamma_{SR}^{(f)} < \gamma_{SR}$. For a larger $N$, less power is transferred to $LP_{01}$ and $\gamma_{SR}^{(f)}$ becomes smaller, resulting in a reduced efficiency $\kappa_0$. Our numerical simulation qualitatively confirms the similar dependence of $\kappa$ and $\kappa_0$ on $N$ (see below and Supplementary Note 5). We find that $\eta$ is about ten times smaller than $\eta^{(bw)}$ that is derived based on the first SR burst. This may be attributed to the fact that the first SR burst does not behave exponentially.

**Discussion**

To understand the subtle mechanism behind the above experimental observations, we need to simulate the collective atom-light interplay under the experimental detection conditions. A similar in-fibre SR model has been studied in a recent theoretical work[39], where the physical system is simplified and only the collective decay of excited atoms, rather than the photon emission, is focused on. Here, we model the practical physical system by employing Maxwell-Bloch equations (MBEs)[2] and derive the in-fibre light field directly. The theoretical model takes into account the effects of the inhomogeneous atomic distribution, the multi-transverse-fibre-mode propagation, and the resonant dipole-dipole interactions[40,41] among atoms in the same lattice site (see Supplementary Note 1-4). The fibre modes are calculated via the full-vector finite element method[42] and the 4th order Runge-Kutta technique is applied to solve MBEs. Choosing fibre modes is crucial for the numerical simulation. Since the SR light is emitted within the cross section of the atomic cloud, the higher-order fibre modes, whose intensity distributions have a central peak with a radius larger than or similar to the radial radius $r_a$ of the atomic cloud, should be considered. We consider up to nine transverse fibre modes, among which the intensity patterns of the modes in Group I (i.e., $LP_{01}$, $LP_{02}$ and $LP_{03}$) are maximized at the central point of the fibre core while the others in Group II (i.e., $LP_{11}^{a,b}$, $LP_{21}^{a,b}$ and $LP_{31}^{a,b}$) do not have central peaks (see Supplementary Note 4), in the simulation. The Gaussian-like fundamental mode $LP_{01}$ has a waist radius $w_0 = 11.8$ μm, at which the intensity drops to $1/e^2$ of the maximum value. The maximum loss (caused by the limited light confinement strength of the cladding design) experienced by the selected modes over the atom-light interaction region is estimated to be lower than $10^{-2}$ dB and is thus negligible. Besides these selected modes, other transverse modes may also propagate inside the actual fibre. However, including more fibre modes requires large amount of computer memory and computational time and may also mismatch the numerical-simulation and experimental results (see below).

Figure 5a displays the simulated SR temporal trace in the same conditions as Fig. 2a. The heterodyne signal $V_{cal}(t)$ is presented in the black colour while the red line shows the envelope of the signal $\tilde{V}_{cal}(t) \propto A'_{SR}(\mathbf{r}, t)$ without the band-pass filtering. It is seen that the ringing behaviour is well replicated. The intervals between two adjacent nodes are consistent with the experimental measurement, proving the validity of simulation. As illustrated by $\tilde{V}_{cal}(t)$, the first node occurs within the pumping process, meaning the SR starts before the pump pulse end. It is also found that the band-pass filter affects the measurement in two aspects: the rapid rising edges of the first two pulses in $\tilde{V}_{\text{cal}}(t)$ are stretched, which explains the fact that the envelope peak corresponding to the pumping pulse in $V_{\text{cal}}(t)$ is lower than that of the first SR burst in $V_{\text{cal}}(t)$; and the variation of $V_{\text{cal}}(t)$ falls behind $\tilde{V}_{\text{cal}}(t)$ by a retarded time of about 0.2 μs. It is worth that the multimode propagation is crucial to this ringing phenomenon. The superposition of multiple transverse modes can sustain an in-fibre beam radius similar to the transverse $1/e$ radius $r_{\text{a}}$ of the atomic cloud, magnifying the collective atom-light interplay (see below and Supplementary Movie 1). We should also note that when extra transverse modes (for example, $LP_{04}$) with a central-peak radius smaller than that of $LP_{03}$ are involved in Group I in the numerical simulation, the separations between inter-adjacent envelop nodes of $V_{\text{cal}}(t)$ are strongly shortened, which is inconsistent with the experimental measurement and manifests the negligible effect of these high-order modes in the atom-light interface.

As we have pointed out above, the multi-transverse-fibre-mode propagation plays an important role in the atom-light coupling. Figure 5b depicts the time-dependent weights of several fibre modes and SR patterns at selected times. The more detailed time evolution of SR pattern can be found in Supplementary Movie 1. The results reveal a unique feature whereby the sequence of the pump absorption and the SR bursts is imprinted in the time evolution of the intensity profile of the fibre guided modes. The initial state of the pump coupling into the fibre is illustrated by the $LP_{01}$-like mode intensity profile at $t = 0$ μs. The doughnut-shaped pattern during the pumping pulse denotes the strong absorption within the cross section of atomic cloud. The SR events are represented by a Gaussian-like pattern with a beam radius smaller than $w_0$. This profile results from a linear combination of the $LP_{01}$, $LP_{02}$ and $LP_{03}$ modes. Hence, the SR in multimode HCPCF can sustain a strong atom-light interplay, which differs from the situation of the atoms coupled to a single-mode fibre or a cavity[4,5]. In addition, around each node $E_{\text{SR}}(\mathbf{r}, t) \approx 0$ of SR, the transverse profile becomes different from the Gaussian-like distribution, indicating the strong influence from the non-central-peak fibre modes in Group II. Only when away from the nodes, the modes in Group I get enhanced prior to the modes in Group II. Moreover, from the numerical simulation one may derive the SR decay rate $\gamma_{\text{bw}}$ corresponding to the first SR burst. Our theoretical result predicts a linear dependence of $\gamma_{\text{bw}}$ on the number of atoms $N$ (see Fig. 5c), $\gamma_{\text{bw}}/\gamma_0 \approx 5.4 \times 10^{-4} N$, consistent with the experimental measurement in Fig. 2b.

Our numerical model also reproduces other experimental results as illustrated in Fig. 6. Performing the Fourier transform on the numerically-simulated amplitude $E_{SR}(\mathbf{r}, t)$, one obtains the SR spectrum $S(\omega) \propto \left| \int_{\tau_p}^{\infty} [\iint E_{SR}(\mathbf{r}, t) d^2\mathbf{r}] e^{i(\omega - \omega_0)t} dt \right|^2$ (see Fig. 6a), from which the frequency shift $\Delta_{SR}$ is read out by deducting the envelope-oscillation frequency introduced by the ringing behaviour. The dependence of $\Delta_{SR}$ on the average on-site atomic density $\rho$ is plotted in Fig. 6b, which agrees with the experimental results. We also theoretically reproduce the counterparts of Fig. 4a and 4b in Fig. 6c and 6d. In comparison, the numerical results of the efficiencies of SR coupling to all fibre modes ($\kappa$) and to the fundamental mode ($\kappa_0$) are both lower than the experimental results. This is mainly because the calculated transverse modes are not exactly the same as those existing in the actual fibre. A slight mismatch can lead to an apparent difference. Furthermore, the insufficient number of the fibre transverse modes in Group II joining in the atom-light interaction may give rise to an erroneous increment of the power absorption within the pumping process, resulting in the reduced $\kappa$ and $\kappa_0$. Nevertheless, the numerical and experimental results still exhibit similar behaviours.

In summary, we have demonstrated the multimode dynamics of collective spontaneous emission based on an engineered in-fibre atom-light-interface architecture. This platform may be applicable to explore more exotic quantum phenomena such as photonic band gaps in 1D ordered atomic structure, quantum jumps of many-body systems with long-range interaction, the open-system Dicke-model phase transition[43], and the competing process between virtual and real photon exchanges for the atom-atom interaction[44]. Confining two atomic clouds at two ends of a long fibre allows the fibre-based remote macro-entanglement and quantum information transmission[45]. Especially, substituting the long-lived $^1S_0 - {}^3P_0$ clock transition for the intercombination transition enables the miniaturization of the physics package of optical lattice clocks[25]. This compact and transportable setup may facilitate the high-level comparison of optical clocks and time keeping[46].

## Methods

**Loading ultracold atoms into HCPCF** Ultracold $^{88}$Sr atoms are prepared via the two-stage MOT[29] and then captured at the antinodes of the optical lattice that is formed by a pair of linearly-polarized counter-propagating $\lambda_L = 813$ nm lasers. Both lattice beams match the fundamental LP$_{01}$ fibre mode with a power of 200 mW. The lattice antinodes are moved along the axial direction (i.e., the z-axis) at a velocity of $v(t) = \delta_L(t)\lambda_L/2$ by introducing time-dependent detuning $|\delta_L(t)| \leq 90$ kHz between two lattice lasers. The atomic cloud is transported by a distance of 3.25 mm within a moving time of about 160 ms at a maximum velocity of 32.5 mm s$^{-1}$ The temperature of atoms is measured to be ~4.0 µK by the Doppler width of the spectrum, which suggests the averaged vibrational occupations of $n_a \sim 0.5$ and $n_r \sim 33$, for the axial and radial motions, respectively.

The atom-light interaction length can be extended via expanding the atomic cloud inside the fibre by shutting off either of lattice lasers to remove the periodic lattice potential. Then, the atomic cloud experiences a Gaussian-like optical potential in the $x - y$ plane and an axial expansion with the velocity satisfying the Maxwell-Boltzmann distribution. After a short duration, the lattice laser, which has been turned off, is switched on again. The atomic cloud with an expanded axial width $l_z$ is recaptured by the lattice potential. During the expanding process, the atom loss is negligible. Figure 1b shows the distribution of the atomic cloud with and without 80-ms-long expansion. We then adiabatically reduced the lattice power to 20 mW, which is one tenth of the initial power, to moderate the density shift and residual light shift for observing SR.

**Mode coupling between HCPCF and SMF** A portion of SR light output from the HCPCF enters the SMF (see Fig. 1a). The optical path between HCPCF and SMF is aligned to maximize the coupling efficiency to the fundamental LP$_{01}$ mode of the HCPCF, which is about 50 %. The Gaussian-like ground mode $\phi_0(\mathbf{r})$ of the SMF approximates as $\phi_0(\mathbf{r}) = \sqrt{2/(\pi \widetilde{w}_0^2)}\, e^{-|\mathbf{r}|^2/\widetilde{w}_0^2}$ with the beam radius of $\widetilde{w}_0 = 0.4 w_0$. The coupling efficiency between LP$_{01}$ and $\phi_0(\mathbf{r})$ is calculated to be $c_{LP_{01}} = |\iint \psi_0^*(\mathbf{r})\phi_0(\mathbf{r})d^2\mathbf{r}|^2 = 0.60$, consistent with the experimental measurement (~50 %). Here $\psi_0(\mathbf{r})$ corresponds to the transverse distribution of LP$_{01}$. Similarly, the coupling efficiencies of selected high-order modes of HCPCF to SMF are given by $c_{LP_{02}} =$ 0.33, $c_{LP_{03}} = 0.05$, and $c_{LP_{11}^{a,b}} = c_{LP_{21}^{a,b}} = c_{LP_{31}^{a,b}} = 0.00$. Thus, the fundamental mode $LP_{01}$ contributes dominantly in energy at the output end of SMF.

**Balanced heterodyne detection** As depicted in Fig. 1c, two Si PIN photodiodes are used to convert the optical-frequency interference down to a radio-frequency (RF) signal $V_{RF}(t) \propto \text{Re}\{[\iint E_{LO}(\mathbf{r})E'_{SR}(\mathbf{r},t)d^2\mathbf{r}]e^{-i\Omega_0 t}\}$, where $\text{Re}\{\ldots\}$ gives the real part. Setting the power of the local optical oscillator $\mathbf{E}_{LO}$ at $P_{LO} \approx 1$ mW, whose shot noise well exceeds the circuit noise, the measurement SNR is approximately given by $\sqrt{2KP_{SR}/(\pi e \Delta f)}$ where $K = 0.42$ A W$^{-1}$ is the photodiodes' responsivity. We further adopt a band-pass filter (see Fig. 1c) with a FWHM bandwidth of $\Delta f = 4$ MHz and a centre frequency of $\Omega_0/2\pi = 50$ MHz to improve the SNR for observing the time-varying radiation amplitude $E_{SR}(\mathbf{r},t)$.

**Deriving radiation frequency shift $\Delta_{SR}$** The 10-times-averaged $V_{RF}(t)$ is demodulated into the in-phase $V_I(t)$ and quadrature-phase $V_Q(t)$ components. We then map the point $V_I(t) + iV_Q(t)$ within the period from 1.48 μs to 8.38 μs into a complex coordinate system (see the purple line in Fig. 3c). Assuming the frequency shift $\Delta_{SR}$ is time independent, the plot can be formulated as $V(t)e^{i\Delta_{SR}t+\theta_0}$, where $V(t) = \sqrt{V_I^2(t) + V_Q^2(t)}$ denotes the distance from the point $V_I(t) + iV_Q(t)$ to the origin of the coordinate and $\theta_0$ is the initial phase, because the phase components are demodulated with the frequency of $\omega_0$ (see Fig. 1d and Fig. 3a). If we rotate the frame around the origin of the coordinate at a frequency of $\omega'$ and with a phase offset of $\theta'$, the plot may be re-expressed as $V(t)e^{i(\Delta_{SR}-\omega')t+(\theta_0-\theta')}$. By assigning $\omega' = \Delta_{SR}$ and $\theta' = \theta_0$, the point moves only on the real axis. To find $\Delta_{SR}$ from two phase components, we prepare an evaluation function that integrates the square of the imaginary part with respect to time and adopt $\omega'$ that minimizes the evaluation function related to $\Delta_{SR}$. The blue line in Fig. 3c shows the plot in the rotating frame that suppresses the imaginary part and $\Delta_{SR}$ is determined to be $\Delta_{SR} \approx -2\pi \times 93(10)$ kHz. Here the uncertainty is estimated by the frequency where the value of the evaluation function is doubled.

## Data availability

All data and computer code supporting the findings of this study are available from the corresponding author on reasonable request.

## Acknowledgements

This work is supported by JST ERATO Grant No. JPMJER1002 10102832 (Japan), by JSPS Grant-in-Aid for Specially Promoted Research Grant No. JP16H06284, and by the Photon Frontier Network Program of the Ministry of Education, Culture, Sports, Science and Technology, Japan (MEXT). FB acknowledges financial supports from Agence Nationale de Recherche (ANR).

## Author contributions

H.K. envisaged and initiated the experiments. H.K. and S.O. designed the apparatus and experiments. S.O. carried out the experiments and analysed the data. S.O. and H.K. discussed experimental results and equally contributed to the experiments. D.Y. developed theoretical models for the experiments. F.B. designed and fabricated the fibre for the experimental requirements and assessed the in-fibre dipole spontaneous emission, and L.V. calculated the fibre modal fields. All authors participated in discussions and the writing of the text.

## Competing interests

The authors declare no competing interests.

## Corresponding author

Correspondence and requests for material should be addressed to H. K. (e-mail: katori@amo.t.u-tokyo.ac.jp).

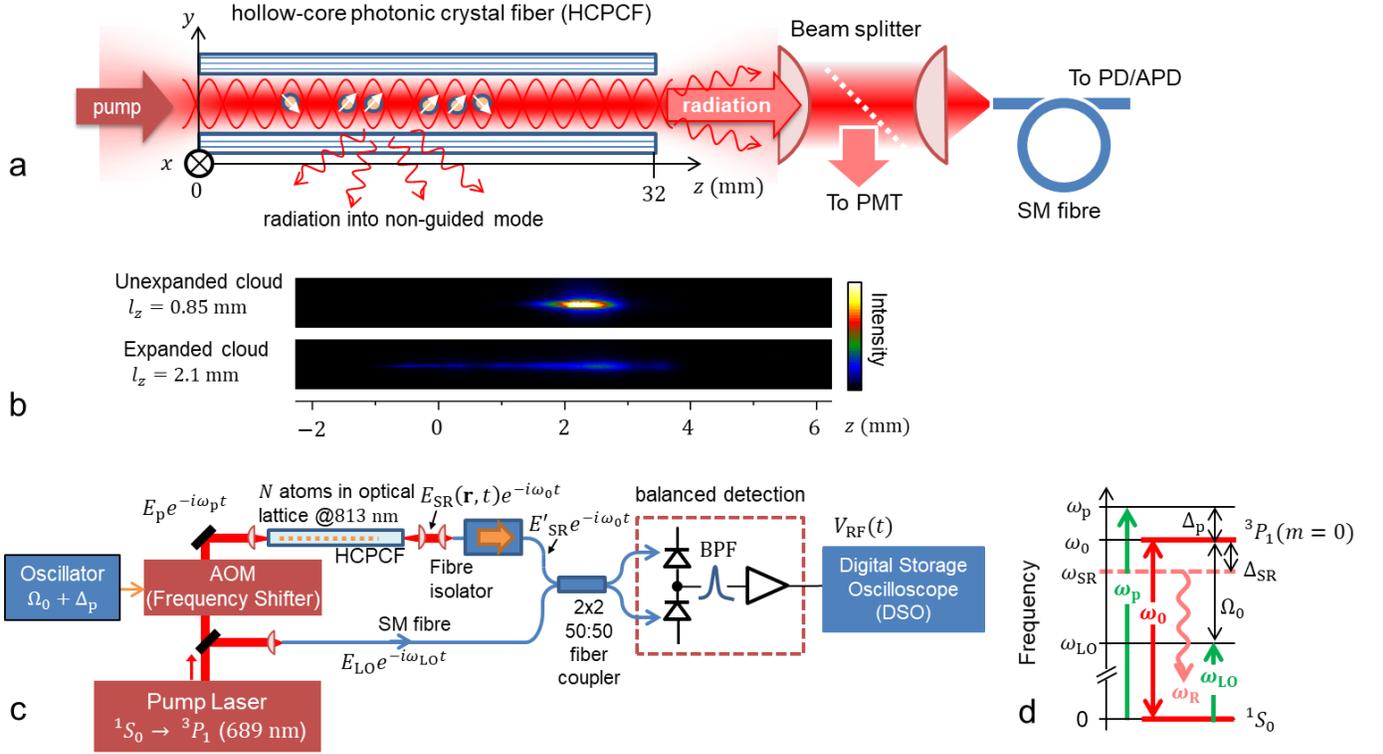

**Figure 1 | Experimental setup. a.** Schematic of experimental setup. Ultracold $^{88}$Sr atoms are loaded from a two-stage magneto-optical trap (MOT) into an optical lattice and then transported inside the hollow-core photonic crystal fibre (HCPCF) at $z = 2$ mm by the moving lattice. The magnetic field $\mathbf{B}_0 = (0.14 \text{ mT}) \hat{\mathbf{e}}_x$ is applied to define the quantization axis of the system. The superradiance (SR) emission strongly couples to the guided modes of the fibre while the spontaneously emitted photons hardly match the fibre modes. The HCPCF guided SR is coupled to a single mode fibre (SMF) that is mainly mode matched to the LP$_{01}$-mode component in SR. A part of the SR output is detected by a photo multiplier tube (PMT) to measure the total intensity of SR including the higher order modes. **b**. Fluorescence images of atoms inside the HCPCF by exciting the $^1S_0 - {}^1P_1$ transition at 461 nm, which are taken by an electron multiplying charge-coupled device (EMCCD) camera through the fibre wall. The FWHM of the atomic distribution is $l_z = 0.87$ mm for the unexpanded cloud and $l_z = 2.1$ mm for the expanded one. **c**. A pump laser at 689 nm with a linewidth less than 1 kHz is split into two paths. One is coupled to a SMF and used as a local oscillator $E_{\text{LO}} \cos \omega_{\text{LO}} t$, while the other is used as the pump light with the amplitude $E_\text{p}$ at the frequency $\omega_\text{p} = \omega_0 + \Delta_\text{p}$ shifted by an acousto-optic modulator (AOM). The SR field from the HCPCF is coupled to the SMF and balanced-heterodyned with the local field.

The resultant signal $V_{\text{RF}}(t) \propto \text{Re}\{[\iint E_{\text{LO}}(\mathbf{r}) E'_{\text{SR}}(\mathbf{r}, t) d^2\mathbf{r}] e^{-i\Omega_0 t}\}$ at around $\Omega_0 = 2\pi \times 50$ MHz is band-pass-filtered (BPF) and recorded by a digital storage oscilloscope (DSO). **d**. Frequency relations. In respect to the $^1S_0 - {^3P_1}(m = 0)$ transition frequency $\omega_0$, the SR, pump-light, and local-oscillator frequencies are given by $\omega_{\text{SR}} = \omega_0 + \Delta_{\text{SR}}$, $\omega_{\text{p}} = \omega_0 + \Delta_{\text{p}}$, and $\omega_{\text{LO}} = \omega_0 - \Omega_0$, respectively.

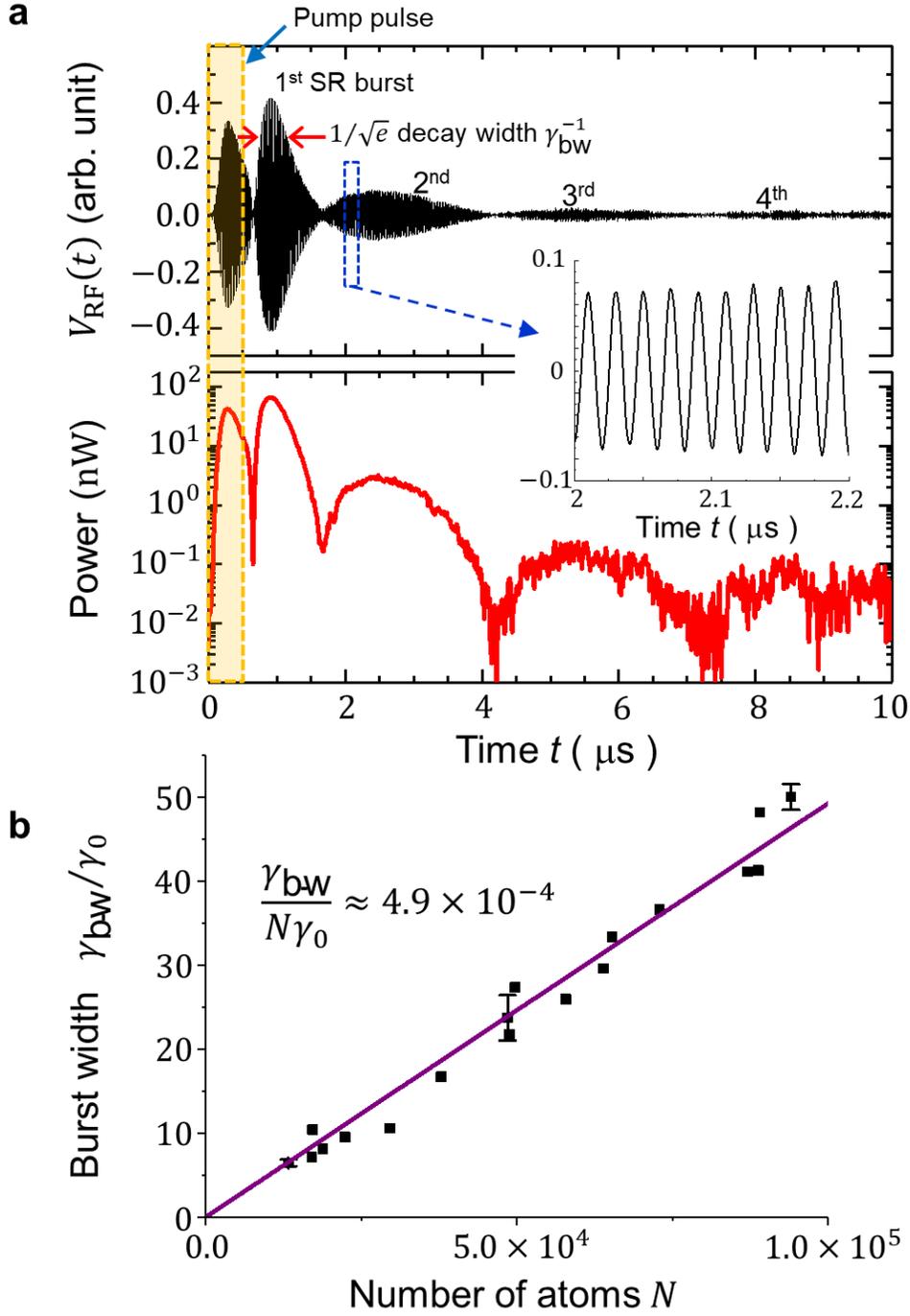

**Figure 2 | Superradiant ringing. a.** Ringing behaviour of the filtered heterodyne signal $V_{RF}(t)$ (black) after applying the resonant ($\Delta_p = 0$) pump pulse with the duration $\tau_p = 0.5$ μs (indicated by a yellow-shaded region) and the amplitude $E_p = \hbar\pi/(D\tau_p)$. $N \approx 9.4 \times 10^4$ atoms are distributed over $l_z = 0.85$ mm along the fibre axis. The first 0.7-μs-long peak corresponds to the

pumping process, which is followed by four superradiance (SR) bursts within 10 μs. The inset shows the zoom up to reveal the carrier oscillation at about $\Omega_0$. The time-dependent SR power proportional to $V_{RF}^2(t)$ is given by the red curve. The SR decay rate $\gamma_{bw}$ of the first SR burst is evaluated by the reciprocal of the temporal width at $1/\sqrt{e}$ of the peak amplitude. For $N \approx 9.4 \times 10^4$, $\gamma_{bw}$ reaches $55\gamma_0$. **b.** Experimentally measured burst width $\gamma_{bw}$ as a function of the number of atoms $N$. The curve fitting gives the linear relation $\gamma_{bw}/\gamma_0 = (4.9 \times 10^{-4}) \times N$, manifesting the SR feature. The error bars, which are given for three representative data, show the standard deviations.

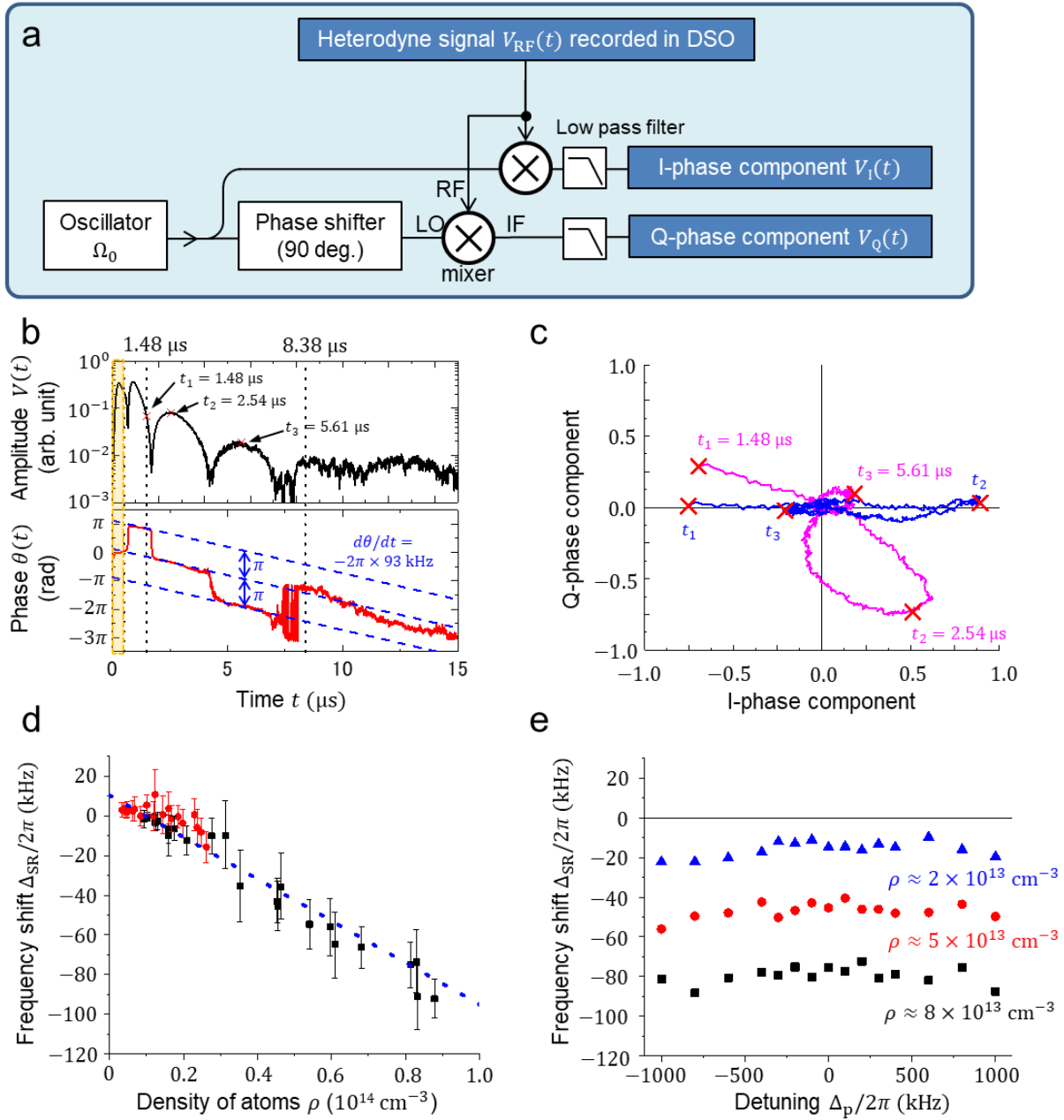

**Figure 3 | Frequency shift of superradiant emission. a.** In-phase $V_I(t)$ and quadrature-phase $V_Q(t)$ components are obtained by demodulating the $V_{RF}(t)$ signal stored in digital storage oscilloscope. **b.** Example of an amplitude $V(t) = \sqrt{V_I(t)^2 + V_Q(t)}$ (black line) and a phase $\theta(t) = \mathrm{ArcTan}[V_Q(t)/V_I(t)]$ (red line) measured for $N \approx 9.4 \times 10^4$. A linear phase shift in time indicates the constant frequency shift of the emitted light. Dashed lines with $d\theta/dt = 2\pi \times 93$ kHz are shown for guides to eyes. **c.** Mapping $V_I(t)$ and $V_Q(t)$ into an I/Q plane. In order to get rid of the influence of the pump field and to ensure enough signal-to-noise ratio, $V_I(t)$ and

$V_Q(t)$ within 1.48 µs $\leq t \leq$ 8.38 µs are adopted to derive the frequency shift to be 93(10) kHz. **d**. Dependence of the frequency shift $\Delta_{SR} = \omega_{SR} - \omega_0$ on the average on-site atomic density $\rho$ under the resonant-pump ($\Delta_p = 0$) condition. Black (red) symbols correspond to unexpanded (expanded) atomic clouds. The error bars are given by the frequency where the value of the evaluation function is doubled. **e.** $\Delta_{SR}$ as a function of the pump-pulse detuning $\Delta_p$ measured for different numbers of atoms $N$. We apply unexpanded atomic clouds with $l_z = 0.85$ mm.

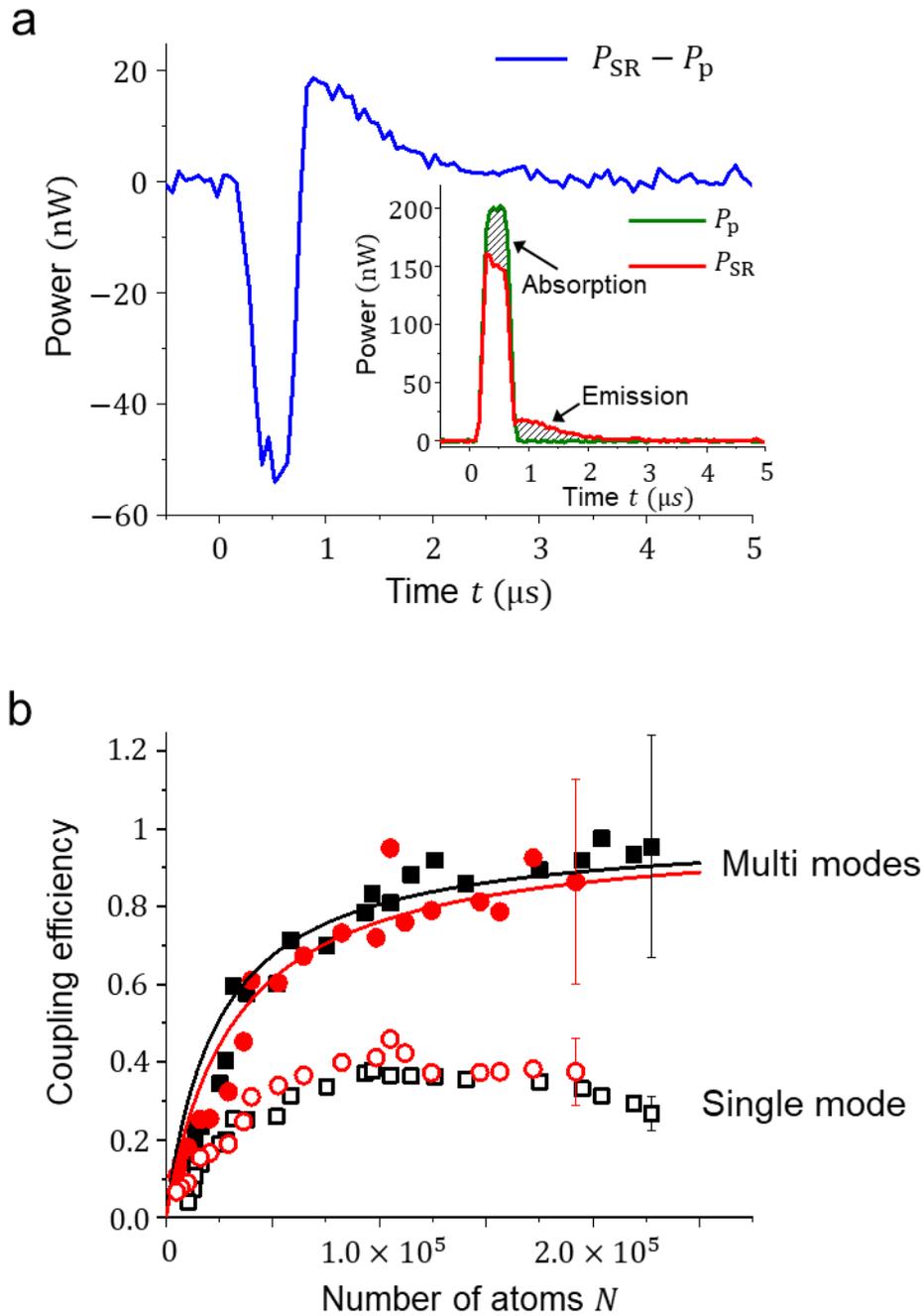

**Figure 4 | Efficiency of superradiance. a**. Time evolution of radiation measured by a photo multiplier tube (PMT), where the pump pulse resonantly ($\Delta_p = 0$) excites the unexpanded atomic cloud with $N \approx 9.4 \times 10^4$ and $l_z = 0.85$ mm. The red (green) curve corresponds to the PMT signal with (without) atoms inside the fibre. The non-overlapping areas within the pumping regimes $t \leq \tau_p$ and $t > \tau_p$ denote the energies absorbed and emitted by the atoms, respectively.

The difference between two curves is given by the blue line. **b**. Total efficiency $\kappa$ of the superradiance (SR) coupling to multiple transverse modes (shown by filled symbols) and efficiency $\kappa_0$ of the SR coupling to the fundamental LP$_{01}$ mode (empty symbols) as a function of the atomic number $N$ for unexpanded ($l_z = 0.85$ mm, shown by black) and expanded ($l_z = 2.1$ mm, shown by red) atomic clouds. Two solid lines give the single-atom cooperativity $\eta = 4.1 \times 10^{-5}$ (filled black), $\eta = 3.2 \times 10^{-5}$ (filled red) with $\chi \approx 1$. The error bars representing the standard deviation are given for four data points.

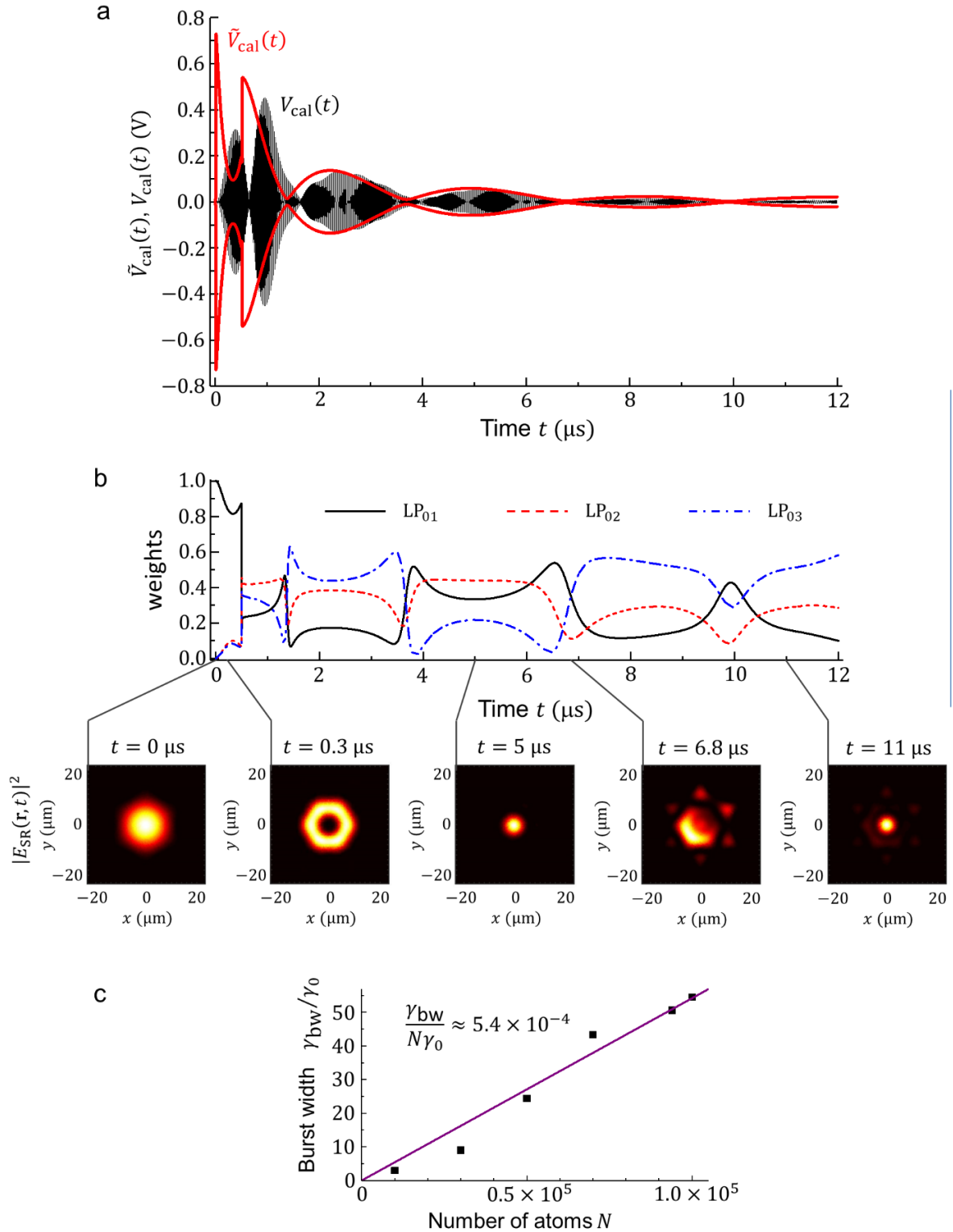

**Figure 5 | Numerical simulation of the ringing behaviour. a.** Numerical result of heterodyne signal $V_{\text{cal}}(t)$ (black) for $N \approx 9.4 \times 10^4$ assuming the resonant ($\Delta_p = 0$) pump pulse with the

duration $\tau_p = 0.5$ μs and the amplitude $E_p = \hbar\pi/(D\tau_p)$, which is corresponding to the experimental measurement. In comparison, the envelope of the raw signal $\tilde{V}_{cal}(t)$ (red) before entering the band-pass filter is also presented. Here, we have used the coupling efficiencies of the fibre $LP_{01}$, $LP_{02}$, and $LP_{03}$ modes to SMF $c_{LP_{01}} = 0.60$, $c_{LP_{02}} = 0.33$, and $c_{LP_{03}} = 0.05$. **b.** Time evolution of fibre-mode weights. The weights of components $LP_{01}$, $LP_{02}$, and $LP_{03}$ in the output SR field $E_{SR}(\mathbf{r}, t)$ vary in time. Examples of SR pattern at several selected times are presented. The pattern deviates from the Gaussian-like distribution at each node of the SR ringing shown in Fig. 5a. Away from ringing nodes, the transverse beam radius of the output SR is about three times smaller than that of the fundamental $LP_{01}$ fibre mode. **c.** Numerical first-SR-burst width $\gamma_{bw}$ vs. the number of atoms $N$. The solid line shows the curve fitting. All system parameters are same to Fig. 5a.

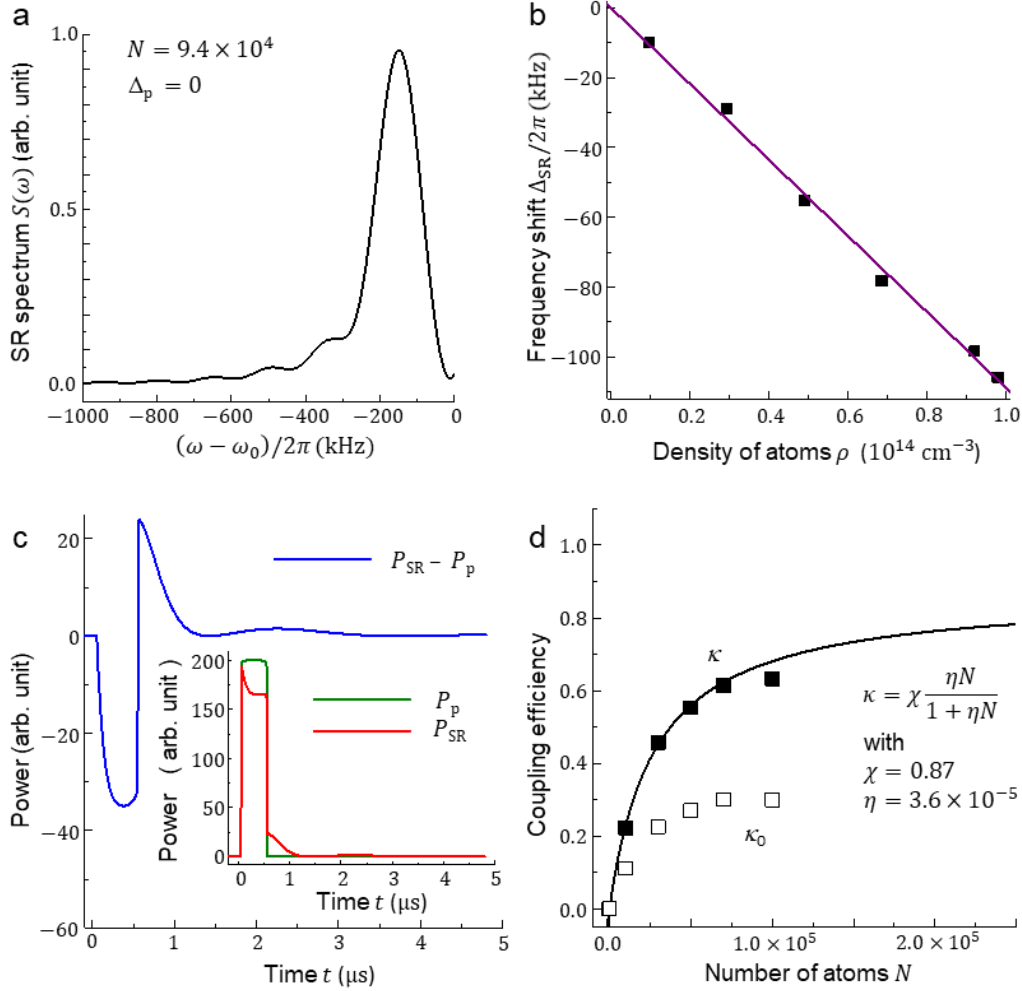

**Figure 6 | Numerical simulation for the resonantly-pumped unexpanded atomic cloud. a**. Superradiance (SR) spectrum $S(\omega)$ with the atom number $N = 9.4 \times 10^4$. The spectrum peak is shifted to the red side of $\omega_0$ by $2\pi \times 152$ kHz. Deducting the envelope-oscillation frequency of the SR field, which is $2\pi \times 54$ kHz, gives $\Delta_{SR} = -2\pi \times 98$ kHz. **b**. Frequency shift $\Delta_{SR}$ vs. the average on-site atomic density $\rho$. The filled black symbols correspond to the numerical results whose curve fitting, i.e., $\Delta_{SR}/\rho = 2\pi \times 1.07 \times 10^{-12}$ kHz cm$^3$, is given by the solid line. **c**. Numerically-derived time-dependent powers of pump beam $P_p(t)$ (green), SR field $P_{SR}(t)$ (red) and their difference $\Delta P(t)$ (blue). The blue line denotes the energies emitted (absorbed) by the atoms when the $\Delta P(t)$ is positive (negative). **d**. Total efficiency $\kappa$ of the SR coupling to multiple transverse modes (filled symbols) and efficiency $\kappa_0$ of the SR coupling to the fundamental LP$_{01}$

mode (empty symbols) as a function of the atomic number $N$. The curve fitting of $\kappa$ leads to $\eta = 3.6 \times 10^{-5}$ and the efficiency $\chi = 0.87$.

## Supplementary Note 1. Physical system

$N$ $^{88}$Sr atoms are confined in a one-dimensional (1D) optical lattice (wavelength $\lambda_\text{L} = 813$ nm) formed inside a hollow-core photonic crystal fibre (HCPCF) along the axial direction (in the $z$-axis). The quantum axis is set by an external magnetic field $\mathbf{B}_0 = (0.14 \text{ mT})\hat{\mathbf{e}}_x$. Here, $\hat{\mathbf{e}}_x$ is the unit vector in the direction of the $x$-axis. The in-fibre atoms are coupled with the superradiant (SR) field

$$\mathbf{E}_\text{SR}(\mathbf{r}, z; t) = (\hat{\mathbf{e}}_x/2)E_\text{SR}(\mathbf{r}, z; t)e^{-i\omega_0 t} + (\hat{\mathbf{e}}_x/2)E_\text{SR}^*(\mathbf{r}, z; t)e^{i\omega_0 t}, \tag{S1}$$

via the $^1S_0 - {}^3P_1(m = 0)$ intercombination transition (frequency $\omega_0 = 2\pi \times 435$ THz, wavelength $\lambda_0 = 2\pi c/\omega_0$ in vacuum, and spontaneous decay rate $\gamma_0 = 2\pi \times 7.5$ kHz). $c$ is the speed of light in vacuum. $\mathbf{r} = x\hat{\mathbf{e}}_x + y\hat{\mathbf{e}}_y$ is the radial coordinate and $\hat{\mathbf{e}}_y$ is the unit vector in the direction of the $y$-axis. $E_\text{SR}(\mathbf{r}, z; t)$ denotes the complex amplitude of SR field. The linear-polarization vector of the optical lattice lasers is chosen to fulfil the magic condition[1] which cancels the light-shift difference between $^1S_0$ and $^3P_1(m = 0)$. The symbol $(u, \mu)$ is used to label the $\mu$-th atom in the $u$-th lattice site.

The characteristic length of the in-fibre confined atomic cloud along the axial direction is $l_z$. In the main text, we only focus on two specific values, i.e., $l_z = 0.87$ mm for the unexpanded cloud and $l_z = 2.1$ mm for the expanded cloud. The corresponding lattice-site numbers $N_\text{L}$ are then given by $N_\text{L} = l_z/(\lambda_\text{L}/2) = 2.1 \times 10^3$ (unexpanded) and $N_\text{L} = 5.2 \times 10^3$ (expanded). Assuming a Gaussian atomic distribution over the optical-lattice region, the number of atoms in the $u$-th lattice site reaches

$$N_u = [N\lambda_\text{L}/(\sqrt{\pi}l_z)] \exp[-(\tilde{z}_u - \tilde{z}_c)^2/(l_z/2)^2], \tag{S2}$$

where $\tilde{z}_u$ denotes the central position of the $u$-th lattice site in the axial direction and $\tilde{z}_c$ is the central position of the atomic cloud along the axial direction. Summing $N_u$ over all lattice sites results in the total number of atoms, i.e., $N = \sum_{u=1}^{N_\text{L}} N_u$. In the $u$-th lattice site, $N_u$ atoms spread within a pancake-shaped potential with the characteristic axial ($z$-axis) width $l_\text{a} = 54$ nm and radial ($x - y$ plane) radius $r_\text{a} = 1.7$ μm and follow the Gaussian distributions

$$[1/(\pi r_\text{a}^2)] \exp\left[-|\mathbf{r}_{(u,\mu)}|^2/r_\text{a}^2\right] \text{ and } [2/(\sqrt{\pi}l_\text{a})] \exp\left[-|z_{(u,\mu)} - \tilde{z}_u|^2/(l_\text{a}/2)^2\right], \tag{S3}$$

in the radial plane and the axial direction, respectively. Here $\mathbf{r}_{(u,\mu)} + z_{(u,\mu)}\hat{\mathbf{e}}_z$ corresponds to the spatial position of the $(u, \mu)$-th atom. $\hat{\mathbf{e}}_z$ is the unit vector in the direction of the $z$-axis.

For the atoms in the same lattice site, the interatomic distance may be shorter than $\lambda_0$, leading to the virtual-photon-mediated resonant dipole-dipole interaction[2]. In contrast, the interaction between two atoms in different lattice sites is weak and negligible because of $(\lambda_L/2) > (\lambda_0/2\pi)$.

## Supplementary Note 2. Equations of motion for atoms

The Hamiltonian describing the coherent atom-light interface and long-range dipolar interaction is written as[2]

$$\hat{H} = \sum_u \sum_\mu^{(u)} \hbar\omega_0 \hat{\sigma}^\dagger_{(u,\mu)} \hat{\sigma}_{(u,\mu)} + \hbar\gamma_0 \sum_u \sum_{\mu_1}^{(u)} \sum_{\mu_2 \neq \mu_1}^{(u)} G_{u,(\mu_1\mu_2)} \hat{\sigma}^\dagger_{(u,\mu_1)} \hat{\sigma}_{(u,\mu_2)}$$
$$+ \sum_u \sum_\mu^{(u)} \left[ -\frac{1}{2} D E_{\text{SR}}(\mathbf{r}_{(u,\mu)}, z_{(u,\mu)}; t) \hat{\sigma}^\dagger_{(u,\mu)} e^{-i\omega_0 t} + \text{h.c.} \right], \quad (S4)$$

where the operator associated with the $(u,\mu)$-th atom is defined as $\hat{\sigma}_{(u,\mu)} = \left( |\,^1S_0\rangle\langle\,^3P_1(m=0)| \right)_{(u,\mu)}$, the dipole moment is $D = (3\pi\hbar\varepsilon_0 c^3 \gamma_0/\omega_0^3)^{1/2}$, and the symbol $\sum_\mu^{(u)} \ldots$ denotes the summation over the atoms in the $u$-th lattice site. $\hbar$ is the reduced Planck's constant and $\varepsilon_0$ is the vacuum permittivity. $E_{\text{SR}}(\mathbf{r}_{(u,\mu)}, z_{(u,\mu)}; t)$ is the SR-field amplitude at the $(u,\mu)$-th atomic position. The parameter

$$G_{u,(\mu_1\mu_2)} = \frac{3}{4}\left[ -(1 - \cos^2\theta_{u,(\mu_1\mu_2)}) \frac{\cos\beta_{u,(\mu_1\mu_2)}}{\beta_{u,(\mu_1\mu_2)}} \right.$$
$$\left. + (1 - 3\cos^2\theta_{u,(\mu_1\mu_2)}) \left( \frac{\sin\beta_{u,(\mu_1\mu_2)}}{\beta^2_{u,(\mu_1\mu_2)}} + \frac{\cos\beta_{u,(\mu_1\mu_2)}}{\beta^3_{u,(\mu_1\mu_2)}} \right) \right], \quad (S5)$$

characterizes the coherent virtual-photon-mediated dipolar interaction between two atoms $(u,\mu_1)$ and $(u,\mu_2)$ in the $\mu$-th lattice site. The normalized interatomic distance is given by $\beta_{u,(\mu_1\mu_2)} = 2\pi|\mathbf{r}_{(u,\mu_1)} - \mathbf{r}_{(u,\mu_2)}|/\lambda_0$ and the angle between the light-induced-dipole vector (along $\hat{\mathbf{e}}_x$) and the $(\mathbf{r}_{(u,\mu_1)} - \mathbf{r}_{(u,\mu_2)})$ vector is $\theta_{u,(\mu_1\mu_2)}$. For $\beta_{u,(\mu_1\mu_2)} < 1$, the $\sim \beta^{-3}_{u,(\mu_1\mu_2)}$ term plays the predominant role, i.e., $G_{u,(\mu_1\mu_2)} \propto \beta^{-3}_{u,(\mu_1\mu_2)}$. The Heisenberg equation for an arbitrary atomic operator $\hat{O}$ is then expressed as

$$\frac{d}{dt}\hat{O} = \sum_u \sum_{\mu_1}^{(u)} \left( i\omega_0 [\hat{\sigma}^\dagger_{(u,\mu_1)} \hat{\sigma}_{(u,\mu_1)}, \hat{O}] + i\gamma_0 \sum_{\mu_2 \neq \mu_1}^{(u)} G_{u,(\mu_1\mu_2)} [\hat{\sigma}^\dagger_{(u,\mu_1)} \hat{\sigma}_{(u,\mu_2)}, \hat{O}] \right)$$
$$+ \sum_u \sum_\mu^{(u)} \left( -i\frac{D}{2\hbar} E_{\text{SR}}(\mathbf{r}_{(u,\mu)}, z_{(u,\mu)}; t) [\hat{\sigma}^\dagger_{(u,\mu)}, \hat{O}] e^{-i\omega_0 t} + \text{h.c.} \right)$$
$$+ \gamma_0 \sum_u \sum_{\mu_1}^{(u)} \sum_{\mu_2}^{(u)} R_{u,(\mu_1\mu_2)} \left( \hat{\sigma}^\dagger_{(u,\mu_1)} \hat{O} \hat{\sigma}_{(u,\mu_2)} \right.$$

$$-\frac{1}{2}\hat{\sigma}^\dagger_{(u,\mu_1)}\hat{\sigma}_{(u,\mu_2)}\hat{O} - \frac{1}{2}\hat{O}\hat{\sigma}^\dagger_{(u,\mu_1)}\hat{\sigma}_{(u,\mu_2)}\Big), \tag{S6}$$

where we have taken into account the dissipative dipolar interatomic coupling characterized by the parameter[2]

$$R_{u,(\mu_1\mu_2)} = \frac{3}{2}\bigg[(1-\cos^2\theta_{u,(\mu_1\mu_2)})\frac{\sin\beta_{u,(\mu_1\mu_2)}}{\beta_{u,(\mu_1\mu_2)}}$$
$$+(1-3\cos^2\theta_{u,(\mu_1\mu_2)})\bigg(\frac{\cos\beta_{u,(\mu_1\mu_2)}}{\beta^2_{u,(\mu_1\mu_2)}} - \frac{\sin\beta_{u,(\mu_1\mu_2)}}{\beta^3_{u,(\mu_1\mu_2)}}\bigg)\bigg], \tag{S7}$$

For $\mathbf{r}_{(u,\mu_1)} = \mathbf{r}_{(u,\mu_2)}$, we have $R_{u,(\mu_1\mu_2)} = 1$.

Based on Supplementary eq. S6, one can derive the Heisenberg equations for the lowering $\hat{\sigma}_{(u,\mu)}$ and population-difference $\hat{w}_{(u,\mu)} = \hat{\sigma}^\dagger_{(u,\mu)}\hat{\sigma}_{(u,\mu)} - \hat{\sigma}_{(u,\mu)}\hat{\sigma}^\dagger_{(u,\mu)}$ operators, i.e.,

$$\frac{d}{dt}\hat{\sigma}_{(u,\mu_1)} = \left(-\frac{\gamma_0}{2} - i\omega_0\right)\hat{\sigma}_{(u,\mu_1)} + \bigg[-i\frac{D}{2\hbar}E_{SR}(\mathbf{r}_{(u,\mu_1)}, z_{(u,\mu_1)}; t)e^{-i\omega_0 t}$$
$$+\gamma_0 \sum_{\mu_2 \neq \mu_1}^{(u)}\left(\frac{R_{u,(\mu_1\mu_2)}}{2} + iG_{u,(\mu_1\mu_2)}\right)\hat{\sigma}_{(u,\mu_2)}\bigg]\hat{w}_{(u,\mu_1)}, \tag{S8a}$$

$$\frac{d}{dt}\hat{w}_{(u,\mu_1)} = -\gamma_0(\hat{w}_{(u,\mu_1)} + 1) - 2\hat{\sigma}^\dagger_{(u,\mu_1)}\bigg[-i\frac{D}{2\hbar}E_{SR}(\mathbf{r}_{(u,\mu_1)}, z_{(u,\mu_1)}; t)e^{-i\omega_0 t}$$
$$+\gamma_0\sum_{\mu_2\neq\mu_1}^{(u)}\left(\frac{R_{u,(\mu_1\mu_2)}}{2} + iG_{u,(\mu_1\mu_2)}\right)\hat{\sigma}_{(u,\mu_2)}\bigg]$$
$$-2\bigg[i\frac{D}{2\hbar}E^*_{SR}(\mathbf{r}_{(u,\mu_1)}, z_{(u,\mu_1)}; t)e^{i\omega_0 t}$$
$$+\gamma_0\sum_{\mu_2\neq\mu_1}^{(u)}\left(\frac{R_{u,(\mu_1\mu_2)}}{2} - iG_{u,(\mu_1\mu_2)}\right)\hat{\sigma}^\dagger_{(u,\mu_2)}\bigg]\hat{\sigma}_{(u,\mu_1)}, \tag{S8b}$$

Defining $\hat{\bar{\sigma}}_{(u,\mu)} = \hat{\sigma}_{(u,\mu)}e^{i\omega_0 t}$ and taking the quantum average, e.g., $\langle\hat{\bar{\sigma}}_{(u,\mu)}\rangle$ and $\langle\hat{w}_{(u,\mu)}\rangle$, we arrive at

$$\frac{d}{dt}\langle\hat{\bar{\sigma}}_{(u,\mu_1)}\rangle = -\frac{\gamma_0}{2}\langle\hat{\bar{\sigma}}_{(u,\mu_1)}\rangle - i\frac{D}{2\hbar}E_{SR}(\mathbf{r}_{(u,\mu_1)}, z_{(u,\mu_1)}; t)\langle\hat{w}_{(u,\mu_1)}\rangle$$
$$+\gamma_0\sum_{\mu_2\neq\mu_1}^{(u)}\left(\frac{R_{u,(\mu_1\mu_2)}}{2} + iG_{u,(\mu_1\mu_2)}\right)\langle\hat{\bar{\sigma}}_{(u,\mu_2)}\hat{w}_{(u,\mu_1)}\rangle, \tag{S9a}$$

$$\frac{d}{dt}\langle\hat{w}_{(u,\mu_1)}\rangle = -\gamma_0(\langle\hat{w}_{(u,\mu_1)}\rangle + 1) + 2i\frac{D}{2\hbar}\bigg[E_{SR}(\mathbf{r}_{(u,\mu_1)}, z_{(u,\mu_1)}; t)\langle\hat{\bar{\sigma}}^\dagger_{(u,\mu_1)}\rangle$$
$$-E^*_{SR}(\mathbf{r}_{(u,\mu_1)}, z_{(u,\mu_1)}; t)\langle\hat{\bar{\sigma}}_{(u,\mu_1)}\rangle\bigg]$$
$$-2\gamma_0\sum_{\mu_2\neq\mu_1}^{(u)}\bigg[\left(\frac{R_{u,(\mu_1\mu_2)}}{2} + iG_{u,(\mu_1\mu_2)}\right)\langle\hat{\bar{\sigma}}^\dagger_{(u,\mu_1)}\hat{\bar{\sigma}}_{(u,\mu_2)}\rangle$$
$$+\left(\frac{R_{u,(\mu_1\mu_2)}}{2} - iG_{u,(\mu_1\mu_2)}\right)\langle\hat{\bar{\sigma}}^\dagger_{(u,\mu_2)}\hat{\bar{\sigma}}_{(u,\mu_1)}\rangle\bigg]. \tag{S9b}$$

We further apply the zero-order approximation, for example,

$$\langle\hat{\bar{\sigma}}_{(u,\mu_1)}(t)\hat{w}_{(u,\mu_2\neq\mu_1)}(t)\rangle \approx \langle\hat{\bar{\sigma}}_{(u,\mu_1)}(t)\rangle\langle\hat{w}_{(u,\mu_2\neq\mu_1)}(t)\rangle,$$

to simplify the mathematical model. The validity of this approximation can be confirmed by comparing the computed SR behavior and the experimental results (see below and main text). The symbols $s_{(u,\mu)}(t)$ and $w_{(u,\mu)}(t)$ are used to respectively replace $\langle \hat{\tilde{\sigma}}_{(u,\mu)}(t) \rangle$ and $\langle \hat{w}_{(u,\mu)}(t) \rangle$. Finally, we obtain the equations of motion for the atoms

$$\frac{d}{dt} s_{(u,\mu_1)} = -\frac{\gamma_0}{2} s_{(u,\mu_1)} - i \left[ \frac{D}{2\hbar} E_{\text{SR}}(\mathbf{r}_{(u,\mu_1)}, \tilde{z}_u; t) \right.$$
$$\left. + i\gamma_0 \sum_{\mu_2 \neq \mu_1}^{(u)} \left( \frac{R_{u,(\mu_1\mu_2)}}{2} + iG_{u,(\mu_1\mu_2)} \right) s_{(u,\mu_2)} \right] w_{(u,\mu_1)}, \quad \text{(S10a)}$$

$$\frac{d}{dt} w_{(u,\mu_1)} = -\gamma_0 \left( w_{(u,\mu_1)} + 1 \right) + 2i \left[ \frac{D}{2\hbar} E_{\text{SR}}(\mathbf{r}_{(u,\mu_1)}, \tilde{z}_u; t) \right.$$
$$\left. + i\gamma_0 \sum_{\mu_2 \neq \mu_1}^{(u)} \left( \frac{R_{u,(\mu_1\mu_2)}}{2} + iG_{u,(\mu_1\mu_2)} \right) s_{(u,\mu_2)} \right] s^*_{(u,\mu_1)}$$
$$- 2i \left[ \frac{D}{2\hbar} E^*_{\text{SR}}(\mathbf{r}_{(u,\mu_1)}, \tilde{z}_u; t) \right.$$
$$\left. - i\gamma_0 \sum_{\mu_2 \neq \mu_1}^{(u)} \left( \frac{R_{u,(\mu_1\mu_2)}}{2} - iG_{u,(\mu_1\mu_2)} \right) s^*_{(u,\mu_2)} \right] s_{(u,\mu_1)}, \quad \text{(S10b)}$$

where we have approximated the amplitude $E_{\text{SR}}(\mathbf{r}_{(u,\mu)}, z_{(u,\mu)}; t)$ at the $(u, \mu)$-th atomic position by the value $E_{\text{SR}}(\mathbf{r}_{(u,\mu)}, \tilde{z}_u; t)$ at the central position of the $u$-th lattice site, i.e., we have assumed that $E_{\text{SR}}(\mathbf{r}_{(u,\mu)}, z_{(u,\mu)}; t)$ does not vary apparently within the length scale of $l_a$.

## Supplementary Note 3. Equation of motion for SR field

We now derive the equation of motion for the light field inside the fibre. Applying Maxwell's equations, one has

$$(\partial_\mathbf{r}^2 - c^{-2}\partial_t^2 + \partial_z^2)\mathbf{E}_{\text{SR}}(\mathbf{r}, z; t) = \mu_0 \partial_t^2 \mathbf{P}(\mathbf{r}, z; t). \quad \text{(S11)}$$

$\mu_0$ is the vacuum permeability. The medium polarizability is given by

$$\mathbf{P}(\mathbf{r}, z; t) = (\hat{\mathbf{e}}_x/2)\chi_e(\mathbf{r})\varepsilon_0 E_{\text{SR}}(\mathbf{r}, z; t)e^{-i\omega_0 t}$$
$$+ \hat{\mathbf{e}}_x D \sum_u \sum_\mu^{(u)} \langle \hat{\sigma}_{(u,\mu)}(t) \rangle \delta(\mathbf{r} - \mathbf{r}_{(u,\mu)}) \delta(z - z_{(u,\mu)}) + \text{c.c.}, \quad \text{(S12)}$$

where $\chi_e(\mathbf{r})$ is the electric susceptibility of the fibre in the absence of lattice-confined atoms. Substituting above expression into Supplementary eq. S11, we are left with

$$[\partial_\mathbf{r}^2 - c^{-2}n^2(\mathbf{r})\partial_t^2 + \partial_z^2] E_{\text{SR}}(\mathbf{r}, z; t)e^{-i\omega_0 t}$$
$$= 2\mu_0 D \partial_t^2 \sum_u \sum_\mu^{(u)} \langle \hat{\tilde{\sigma}}_{(u,\mu)}(t) \rangle e^{-i\omega_0 t} \delta(\mathbf{r} - \mathbf{r}_{(u,\mu)}) \delta(z - z_{(u,\mu)}). \quad \text{(S13)}$$

Here $n(\mathbf{r}) = \sqrt{1 + \chi_e(\mathbf{r})}$ denotes the refractive-index profile of the fibre's cross-section.

In the absence of atoms, one obtains a time-independent equation from (S13),

$$[\partial_{\mathbf{r}}^2 + n^2(\mathbf{r})\omega_0^2/c^2 + \partial_z^2]E(\mathbf{r},z) = 0, \tag{S14}$$

for a general in-fibre light amplitude $E(\mathbf{r},z)$. Following the separation of variables procedure, Supplementary eq. S14 leads to the $m$-th transverse eigenmode $\psi_m(\mathbf{r})$ with the corresponding effective refractive index $n_m^{\text{eff}}$, i.e.,

$$[\partial_{\mathbf{r}}^2 + n^2(\mathbf{r})\omega_0^2/c^2]\psi_m(\mathbf{r})e^{in_m^{\text{eff}}\omega_0 z/c} = \left(n_m^{\text{eff}}\omega_0/c\right)^2\psi_m(\mathbf{r})e^{in_m^{\text{eff}}\omega_0 z/c}, \tag{S15a}$$

$$\partial_z^2\psi_m(\mathbf{r})e^{in_m^{\text{eff}}\omega_0 z/c} = -\left(n_m^{\text{eff}}\omega_0/c\right)^2\psi_m(\mathbf{r})e^{in_m^{\text{eff}}\omega_0 z/c}. \tag{S15b}$$

The normalization condition is given by $\iint \psi_m^*(\mathbf{r})\psi_{m'}(\mathbf{r})d^2\mathbf{r} = \delta_{m,m'}$. $\psi_0(\mathbf{r})$ corresponds to the Gaussian-like fundamental HE$_{11}$ mode. Based on $\psi_m(\mathbf{r})$, the SR amplitude $E_{\text{SR}}(\mathbf{r},z;t)$ may be rewritten in a superposition form

$$E_{\text{SR}}(\mathbf{r},z;t) = \sum_m f_m(z,t)\psi_m(\mathbf{r})e^{in_m^{\text{eff}}\omega_0 z/c}. \tag{S16}$$

$|f_m(z,t)|^2$ correspond to the mode weights. Substituting Supplementary eq. S16 into Supplementary eq. S13 and using the slowly varying envelope approximation, $|\partial_t f_m(z,t)| \ll \omega_0|f_m(z,t)|$ and $|\partial_z f_m(z,t)| \ll \left(n_m^{\text{eff}}\omega_0/c\right)|f_m(z,t)|$, we arrive at

$$\sum_m \left(n_m^{\text{eff}} + 1\right)[(\partial_z + c^{-1}\partial_t)f_m(z,t)]\psi_m(\mathbf{r})e^{in_m^{\text{eff}}\omega_0 z/c}$$
$$= i\frac{2D\omega_0}{\varepsilon_0 c}\sum_u \sum_\mu^{(u)} \langle\hat{\bar{\sigma}}_{(u,\mu)}(t)\rangle\delta(\mathbf{r} - \mathbf{r}_{(u,\mu)})\delta(z - z_{(u,\mu)}), \tag{S17}$$

which further leads to

$$(\partial_z + c^{-1}\partial_t)f_m(z,t)$$
$$= i\frac{2D\omega_0}{\varepsilon_0 c(n_m^{\text{eff}}+1)}\sum_u \sum_\mu^{(u)} \langle\hat{\bar{\sigma}}_{(u,\mu)}(t)\rangle\psi_m^*(\mathbf{r}_{(u,\mu)})e^{-in_m^{\text{eff}}\omega_0 z_{(u,\mu)}/c}\delta(z - z_{(u,\mu)}). \tag{S18}$$

We integrate Supplementary eq. S18 around the central position $\tilde{z}_u$ of the $u$-th lattice site within the range of $\tilde{z}_u - \lambda_\text{L}/4 < z < \tilde{z}_u + \lambda_\text{L}/4$,

$$\int_{\tilde{z}_u-\lambda_\text{L}/4}^{\tilde{z}_u+\lambda_\text{L}/4}(\partial_z + c^{-1}\partial_t)f_m(z,t)dz = f_m(\tilde{z}_u + \lambda_\text{L}/4, t) - f_m(\tilde{z}_u - \lambda_\text{L}/4, t)$$
$$+ c^{-1}\partial_t \int_{\tilde{z}_u-\lambda_\text{L}/4}^{\tilde{z}_u+\lambda_\text{L}/4} f_m(z,t)dz. \tag{S19}$$

and obtain

$$f_m(\tilde{z}_u + \lambda_\text{L}/4, t) = f_m(\tilde{z}_u - \lambda_\text{L}/4, t)$$
$$+ i\frac{2D\omega_0}{\varepsilon_0 c(n_m^{\text{eff}}+1)}\sum_\mu^{(u)} s_{(u,\mu)}(t)\psi_m^*(\mathbf{r}_{(u,\mu)})e^{-in_m^{\text{eff}}\omega_0 z_{(u,\mu)}/c}. \tag{S20}$$

In deriving Supplementary eq. S20 we have substituted $s_{(u,\mu)}(t)$ for $\langle\hat{\bar{\sigma}}_{(u,\mu)}(t)\rangle$ and omitted the term $c^{-1}\partial_t \int_{\tilde{z}_u-\lambda_\text{L}/4}^{\tilde{z}_u+\lambda_\text{L}/4} f_m(z,t)dz$ in Supplementary eq. S19 since its effect is negligible in the slowly varying envelope approximation. Further, the $e^{-in_m^{\text{eff}}\omega_0 z_{(u,\mu)}/c}$ term in Supplementary eq. S20 may be approximated as $e^{-in_m^{\text{eff}}\omega_0 \tilde{z}_u/c}$ due to the fact $\omega_0 l_a/c \ll 1$. Defining $\tilde{f}_{u,m}(t) = f_m(\tilde{z}_u - \lambda_L/4, t)e^{in_m^{\text{eff}}\omega_0(\tilde{z}_u - \lambda_L/4)/c}$, one has

$$\tilde{f}_{u+1,m}(t)e^{-in_m^{\text{eff}}\omega_0(\lambda_L/4)/c} = \tilde{f}_{u,m}(t)e^{in_m^{\text{eff}}\omega_0(\lambda_L/4)/c}$$
$$+ i\frac{2D\omega_0}{\varepsilon_0 c(n_m^{\text{eff}}+1)}\sum_\mu^{(u)} s_{(u,\mu)}(t)\psi_m^*(\mathbf{r}_{(u,\mu)}), \tag{S21}$$

where we have used the relation $\tilde{z}_{u+1} - \tilde{z}_u = \lambda_L/2$. From Supplementary eq. S21 one may solve the time-evolved SR amplitude $E_{\text{SR}}(\mathbf{r}, \tilde{z}_u - \lambda_L/4; t) = \sum_m \tilde{f}_{u,m}(t)\psi_m(\mathbf{r})$ at the spot of $[\mathbf{r} + (\tilde{z}_u - \lambda_L/4)\hat{\mathbf{e}}_z]$.

$E_{\text{SR}}(\mathbf{r}, \tilde{z}_{u=1} - \lambda_L/4; t)$ ($u = 1$ denotes the first lattice site) is determined by the input pump-field amplitude $E_p(\mathbf{r}, t)$ which is linearly polarized in the $x$-direction. In experiment, $E_p(\mathbf{r}, t)$ is a $\pi$-pulse with a duration $\tau_p = 500$ ns. Applying the slowly varying envelope approximation, the SR amplitude at the central position of the $u$-th lattice site may be approximated as $E_{\text{SR}}(\mathbf{r}, \tilde{z}_u; t) = \sum_m \tilde{f}_{u,m}^{(1/2)}\psi_m(\mathbf{r})e^{in_m^{\text{eff}}\omega_0 \tilde{z}_u/c}$ with

$$2\tilde{f}_{u,m}^{(1/2)} = \tilde{f}_{u,m}(t)e^{in_m^{\text{eff}}\omega_0(\lambda_L/4)/c} + \tilde{f}_{u+1,m}(t)e^{-in_m^{\text{eff}}\omega_0(\lambda_L/4)/c}. \tag{S22}$$

$E_{\text{SR}}(\mathbf{r}, \tilde{z}_{u=N_L+1} - \lambda_L/4; t)$ corresponds to the SR field output from the fibre. For simplicity, in the following we use the symbol $E_{\text{SR}}(\mathbf{r}, t)$ to denote the amplitude of the output SR.

## Supplementary Note 4. Multiple transverse modes

The HCPCF used in experiment supports multimode propagation. For the convenience here we apply the scalar wave approximation to the linearly polarized mode classification, $LP_{l,\mu}$. $l$ and $\mu$ are the indices corresponding to the azimuthal and radial field variation, respectively. Thus, the $HE_{11}$ mode is labelled as $LP_{01}$. The fibre guides via Inhibited Coupling mechanism[3]. Consequently, its modes are leaky and suffer of confinement loss. Confinement loss coefficient quantifies the fraction of power lost by the mode due to the leakage in the fibre cladding per unit of length. Supplementary Figure 1 lists the numerically simulated (using a full-vector finite element method[3]) mode intensity profiles of the first nine lowest-loss guided modes of the HCPCF along with their respective effective refractive indices $n_{m=0,1,\ldots,8}^{\text{eff}}$ and confinement loss coefficients $\alpha_{m=0,1,\ldots,8}$. The results show that the $LP_{01}$ mode has the lowest confinement loss of ~0.01 dB m$^{-1}$ at 813 nm while the $LP_{02}$ mode suffers the highest loss of 4 dB m$^{-1}$. The fundamental $LP_{01}$ mode has a beam radius $w_0 = 11.8$ μm, showing a good matching with the pump field, i.e., $E_p(\mathbf{r},t) \propto \psi_0(\mathbf{r})e^{-i\Delta_p t}$ ($\Delta_p = \omega_p - \omega_0$ is the detuning and $\omega_p$ is the pump-field frequency).

To keep our computation load to a reasonable level for solving our equations of motion, we limit the number of transverse modes to a maximum of 9, i.e., to the ones shown in Supplementary Fig. 1. It is noteworthy that when choosing the most dominant modes in the SR dynamics, both the loss of the fibre modes and their coupling strength with the atomic cloud have been taken into account: (i) Indeed, for the fibre modes with the comparable confinement losses, the modes, whose intensity distributions show a maximum at the centre of the fibre cross section (i.e., $LP_{0,\mu}$), have the stronger coupling strength with the atomic cloud than those whose intensity distributions have a zero at the centre of the fibre core (i.e., $LP_{l,\mu}$ with $l > 0$); and (ii) The confinement loss increases with high orders $l$ and $\mu$. Consequently, the atoms hardly interact with the modes, whose intensity distributions peak at the central point of the fibre core but have the peak diameters smaller than $2r_a$ or do not have central peaks at all.

## Supplementary Note 5. Numerical simulation

The time evolution of the whole system can be numerically simulated based on Supplementary eq. S10a, S10b and S21 via the fourth-order Runge-Kutta technique. The spatial positions of $N$ atoms are randomly generated according to $[(N\lambda_L)/(\sqrt{\pi}l_z)]e^{-(\tilde{z}_u - \tilde{z}_c)^2/(l_z/2)^2}$ and the Gaussian distributions of $(\pi r_a^2)^{-1}\exp[-|\mathbf{r}_{(u,\mu)}|^2/r_a^2]$ in the radial plane and $[2/(\sqrt{\pi}l_a)]\exp[-|z_{(u,\mu)} - \tilde{z}_u|^2/(l_a/2)^2]$ in the axial direction (see Supplementary Fig. 2a and 2b). For the $u$-th lattice site, the average on-site atomic density is

$$\rho_u = N_u/(\pi r_a^2 l_a) = \rho(2/\sqrt{\pi})\exp[-(\tilde{z}_u - \tilde{z}_c)^2/(l_z/2)^2], \tag{S23}$$

where we have defined the atomic density $\rho = N_a/(\pi r_a^2 l_a)$ averaged over the whole lattice region with the averaged on-site atomic number $N_a = N/N_L$.

Our numerical simulation well reproduces the ringing behaviour observed in experiment. The raw heterodyne signal $\tilde{V}_{cal}(t)$ described in Fig. 5a of the main text is given by $\tilde{V}_{cal}(t) \propto \mathrm{Re}\left[e^{-i\Omega_0 t}\iint E_{LO}(\mathbf{r})E'_{SR}(\mathbf{r},t)d^2\mathbf{r}\right]$, where Re[...] is the real part and $\Omega_0 = \omega_0 - \omega_{LO}$. $E_{LO}(\mathbf{r})$ and $\omega_{LO}$ are the amplitude and frequency of the local optical oscillator. The amplitude $E'_{SR}(\mathbf{r},t)$ corresponds to the SR field propagating in the single-mode fibre (shown in Fig. 1c of the main text) and is given by $E'_{SR}(\mathbf{r},t) = [\iint \phi_0^*(\mathbf{r}) E_{SR}(\mathbf{r},t) d^2\mathbf{r}] \phi_0(\mathbf{r})$. Here $\phi_0(\mathbf{r}) = \sqrt{2/(\pi \widetilde{w}_0^2)}\, e^{-|\mathbf{r}|^2/\widetilde{w}_0^2}$ is the Gaussian-like ground mode of the single-mode fibre with the beam radius of $\widetilde{w}_0 = 0.4 w_0$. The band-pass filter removes the unwanted rapid varying component in $\widetilde{V}_{cal}(t)$, leading to the amplitude $V_{cal}(t)$. Figure 5a of the main text shows the numerical result $V_{cal}(t)$ that corresponds to the experimental result $V_{RF}(t)$ shown in Fig. 2a of the main text. It is seen that the theoretical result is consistent with the experimental measurement, proving the validity of the approximation $\langle \hat{\bar{\sigma}}_{(u,\mu_1)}(t) \hat{w}_{(u,\mu_2 \neq \mu_1)}(t) \rangle \approx \langle \hat{\bar{\sigma}}_{(u,\mu_1)}(t) \rangle \langle \hat{w}_{(u,\mu_2 \neq \mu_1)}(t) \rangle$, i.e., the effect of quantum correlation on the SR behavior is negligible. Supplementary Movie 1 also displays an example time evolution of SR for the unexpanded atomic cloud with $N = 9.4 \times 10^4$. The patterns of SR intensity ($\propto |E_{SR}(\mathbf{r},t)|^2$) at several selected times are displayed in Fig. 5b of the main text. The dependence of the decay rate $\gamma_{bw}$ of the first SR burst on the atom number $N$ can be also derived from our simulation as shown in Fig. 5c.

One may further extract the frequency shift $\Delta_{SR} = \omega_{SR} - \omega_0$ of the SR central frequency $\omega_{SR}$ relative to the atomic transition $\omega_0$ from the SR spectrum $S(\omega) \propto \left| \int_{\tau_p}^{\infty} [\iint E_{SR}(\mathbf{r},t) d^2\mathbf{r}] e^{i(\omega-\omega_0)t} dt \right|^2$. The resulting $S(\omega)$ is depicted in Fig. 6a of the main text, where the spectrum peak is red shifted from the atomic resonance $\omega_0$ by $\sim 2\pi \times 152$ kHz. However, this frequency shift also includes the offset frequency $\sim 2\pi \times 54$ kHz of the envelope varying. Consequently, the SR (carrier) frequency shift $\Delta_{SR}$ is equal to $-2\pi \times 98$ kHz, which reasonably agrees with the experimental results in Fig. 3d of the main text. The numerically-derived dependence of $\Delta_{SR}$ on the atomic density $\rho$ is shown in Fig. 6b of the main text.

From the simulation results, one can further compute the superradiance efficiency. The total power of the pump field is given by $P_p(t) \propto \iint |E_p(\mathbf{r},t)|^2 d\mathbf{r}$ while the total power of the SR field output from the fibre is $P_{SR}(t) \propto \iint |E_{SR}(\mathbf{r},t)|^2 d\mathbf{r}$. Figure 6c of the main text shows the time-dependent $P_p(t)$ and $P_{SR}(t)$ for the unexpanded atomic cloud ($l_z = 0.87$ mm) with $N = 9.4 \times 10^4$. Within the $\pi$-pulse period, $P_p(t)$ is higher than $P_{SR}(t)$ and the difference $P_{ab} = \int_0^{\tau_p} (P_p(t) - P_{SR}(t)) dt$ corresponds to the energy absorbed by the atomic cloud. In contrast, $P_{em} = \int_{\tau_p}^{\infty} P_{SR}(t) dt$ denotes the SR-field energy emitted by the excited atoms. We should point out that SR may start before $\tau_p$. The SR efficiency $\kappa$ defined in the main text is then given by $\kappa = P_{\text{em}}/P_{\text{ab}}$. Figure 6d of the main text plots the dependence of $\kappa$ on the atomic number $N$ for the unexpanded cloud (i.e., $l_z = 0.87$ mm), corresponding to Fig. 4b of the main text. We see that $\kappa$ goes up as $N$ is increased and is saturated eventually. Curve fitting leads to $\kappa = \chi(\eta N)/(1 + \eta N)$ with the coupling coefficient $\chi = 0.87$ and the single-atom cooperativity parameter $\eta = 3.6 \times 10^{-5}$. We find that $\kappa$ presented in Fig. 6d of the main text is lower than that of Fig. 4b in the main text. This is mainly attributed to the difference between the calculated transverse fibre eigenmodes $\psi_{m=0,\ldots,8}(\mathbf{r})$ and those propagating in the real fibre. In addition, the insufficient number of the fibre modes, whose intensity profiles do not peak at the centre of the fibre core, joining in the atom-light interaction may also reduce the numerically-simulated efficiency $\kappa$.

We also calculate the fundamental-mode radiation efficiency $\kappa_0$, where the estimated coupling efficiencies for different HC-PCF eigenmodes to the single-mode fibre are: 0.60 for $|\psi_0(\mathbf{r})|^2$, 0.33 for $|\psi_5(\mathbf{r})|^2$, 0.05 for $|\psi_5(\mathbf{r})|^2$, and 0.00 for others. Figure 6d of the main text also depicts the numerical results of $\kappa_0$ corresponding to Fig. 4b of the main text.

## Supplementary References

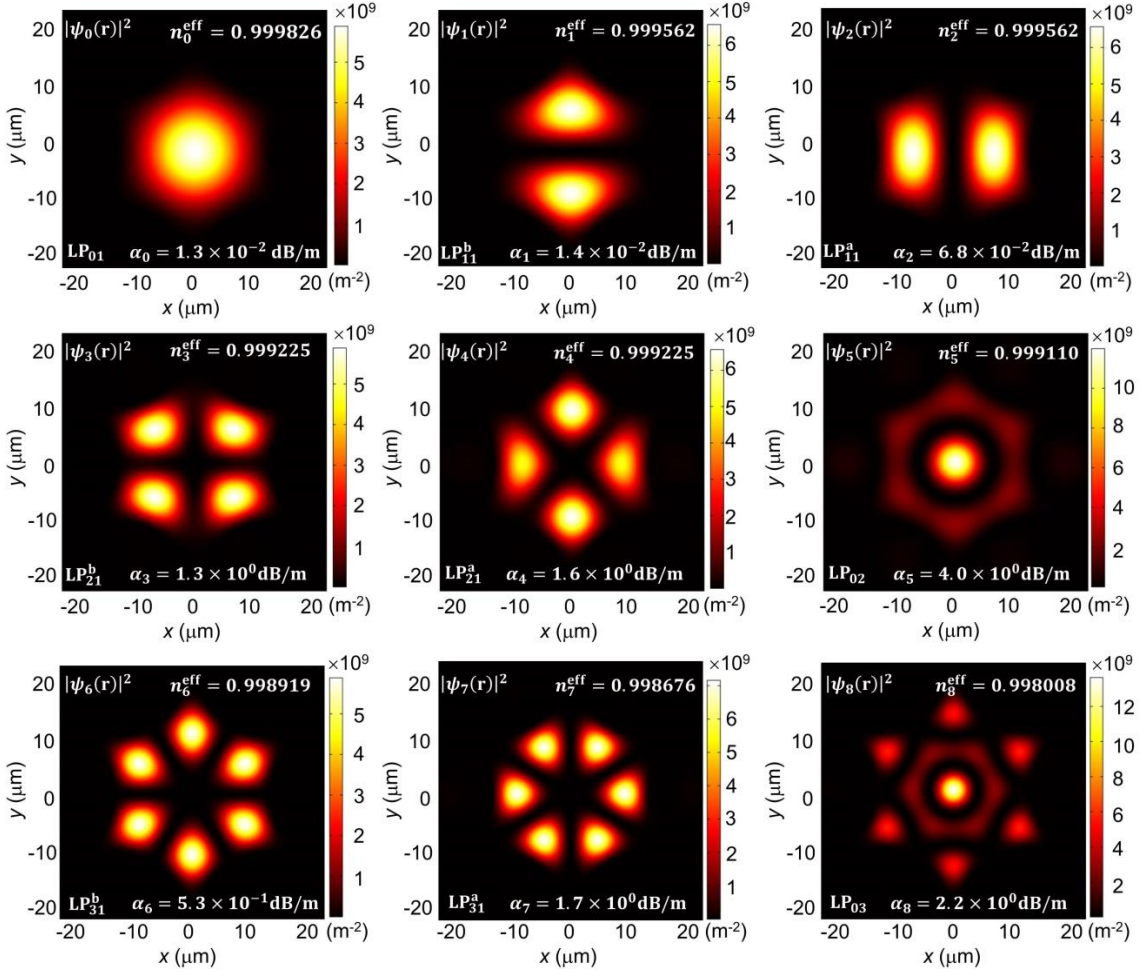

**Supplementary Figure 1 | Transverse Fibre modes.** Intensity distribution of different $\psi_{m=0,1,\ldots,8}(\mathbf{r})$ fibre modes and their effective refractive indices $n^{\text{eff}}_{m=0,1,\ldots,8}$ and confinement loss coefficients $\alpha_{m=0,1,\ldots,8}$ derived by using the full-vector finite element method[3]. The intensity profiles of $\psi_0(\mathbf{r})$, $\psi_5(\mathbf{r})$ and $\psi_8(\mathbf{r})$ own the central peaks while that of others minimize at the central point of fibre cross section.

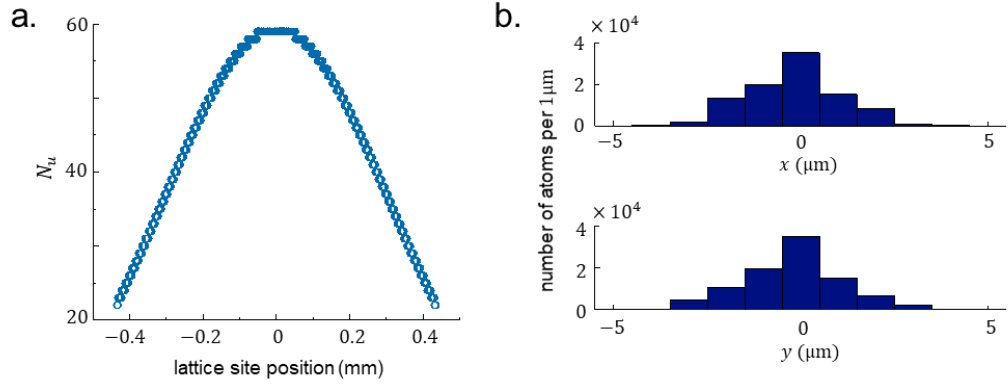

**Supplementary Figure 2** | $N = 9.4 \times 10^4$ atoms inside lattice region. (a) Number $N_u$ of atoms in different lattice sites. The characteristic width of the atomic cloud is $l_z = 0.87$ mm. (b) Histograms of atomic distributions along $x$- and $y$-directions. The characteristic radial radius of the atomic cloud is $r_a = 1.7$ μm.